\documentclass[conference]{IEEEtran}
\pdfoutput=1
\IEEEoverridecommandlockouts
\usepackage{cite}
\usepackage{amsmath,amssymb,amsfonts}
\usepackage{booktabs}

\newtheorem{definition}{Definition}

\newtheorem{proposition}{Proposition}

\usepackage{algorithmic}
\usepackage{graphicx}
\usepackage{subfig}
\usepackage{epstopdf}
\usepackage{textcomp}
\usepackage{xcolor}

\def\BibTeX{{\rm B\kern-.05em{\sc i\kern-.025em b}\kern-.08em
    T\kern-.1667em\lower.7ex\hbox{E}\kern-.125emX}}
\begin{document}

\title{ULA Fitting for Sparse Array Design 
\\
\author{Wanlu Shi, Sergiy A. Vorobyov, \IEEEmembership{Fellow, IEEE},  and Yingsong Li, \IEEEmembership{Senior Member, IEEE}}
\thanks{This work was supported in parts by the China scholarship council under Grant 201906680047, the Ph.D. Student Research and Innovation Fund of the Fundamental Research Funds for the Central Universities under Grant 3072019GIP0808, and Academy of Finland under Grant 319822. Some preliminary results that leaded to this paper have been submitted to {\it IEEE Int. Conf. Acoust., Speech, Signal Process. 2021}, Toronto, Canada.}
\thanks{Wanlu Shi (e-mail: shiwanlu@hrbeu.edu.cn) is with the College of Information and Communication Engineering, Harbin Engineering University, Harbin 150001, China. He was also a visiting student at the Department of Signal Processing and Acoustics, Aalto University, Espoo, Finland. Sergiy A. Vorobyov (email: sergiy.vorobyov@aalto.fi) is with the Department of Signal Processing and Acoustics, Aalto University, Espoo, Finland. Yingsong Li (e-mail: liyingsong@ieee.org) is with the College of Information and Communication Engineering, Harbin Engineering University, Harbin 150001, China. {\it (Corresponding author: Yingsong Li)}}
}

\maketitle
\begin{abstract}
Sparse array (SA) geometries, such as coprime and nested arrays, can be regarded as a concatenation of two uniform linear arrays (ULAs). Such arrays lead to a significant increase of the number of degrees of freedom (DOF) when the second-order information is utilized, i.e., they provide long virtual difference coarray (DCA). Thus, the idea of this paper is based on the observation that SAs can be fitted through concatenation of sub-ULAs. A corresponding SA design principle, called ULA fitting, is then proposed. It aims to design SAs from sub-ULAs. Towards this goal, a polynomial model for arrays is used, and based on it, a DCA structure is analyzed if SA is composed of multiple sub-ULAs. SA design with low mutual coupling is considered. ULA fitting enables to transfer the SA design requirements, such as hole free, low mutual coupling and other requirements, into pseudo polynomial equation, and hence, find particular solutions. We mainly focus on designing SAs with low mutual coupling and large uniform DOF. Two examples of SAs with closed-form expressions are then developed based on ULA fitting. Numerical experiments verify the superiority of the proposed SAs in the presence of heavy mutual coupling.
\end{abstract}


\section{Introduction}
Array signal processing has a pivotal role in communications~\cite{communication}, radar~\cite{radar}, sonar~\cite{sonar}, and other applications due to the abilities of spatial selectivity, suppressing interferences and increasing signal quality. Based on the conventional uniform linear array (ULA) with maximum $\lambda/2$ interspace between antenna elements (here  $\lambda$ is the signal wavelength), many algorithms have been developed to perform beamforming and direction-of-arrival (DOA) estimation~\cite{application_array, music, optimum_array_processing, Vorob1, Vorob2, L1_LMS, fourth_order, KR_subspace}. However, the degrees of freedom (DOF) of ULA is proportional to the number of sensors. Utilizing traditional subspace based techniques, such as MUSIC~\cite{music}, a ULA with $N$ sensors can resolve up to $N-1$ sources. Besides, although ULAs with maximum $\lambda/2$ inter-element spacing successfully avoid spatial aliasing, they may suffer from mutual coupling effect.

The problem of detecting more sources than sensors has long been a problem of great interest in a wide range of fields. To further increase the DOF, the concept of difference coarray (DCA) has been developed. Using DCA, an $N$-sensor sparse array (SA) can have up to $\mathcal{O}(N^2)$ virtual consecutive sensors~\cite{coarray}. The well known minimum redundancy array (MRA) and minimum hole array (MHA), which are in fact SA structures with difference coarrays, provide larger DOF~\cite{MRA,MHA}. Such SAs, however, have failed to lead to a simple closed-form expression for representing them, which is a disadvantage for practical applications.

\par
Recent progress in nested~\cite{NESTED} and coprime arrays~\cite{coprimeC,COPRIME} have indicated the need for SA structures with closed-form expression and long DCAs. Inspired by this, a number of SAs have been introduced and analyzed~\cite{CADIS, hole_filling, Extended_coprime, improved_nested, 2D_nested1, 2D_nested2, multi_nested}. Further, the development of super nested array (SNA)~\cite{SNAC1,SNAC2,SNA1,SNA2} has led to a renewed interest in designing SAs with both long DCA and low mutual coupling. In~\cite{SNA1} and~\cite{SNA2}, the authors have proposed the SNA geometry and emphasized the influence of weight function $w(n)$ (where $w(n)$ refers to the number of sensor pairs with interval $n\lambda/2$) on mutual coupling. Qualitatively, the first three weights of DCA, i.e., $w(1)$, $w(2)$, and $w(3)$, dominate the mutual coupling. SNA achieves a reduced mutual coupling and equivalent DOF compared to nested arrays. In~\cite{ANA}, augmented nested array (ANA) has been developed by rearranging the sensors with small separations in nested array leading to an increased DOF and decreased mutual coupling. Essentially, SNA and ANA are originated from the conventional nested array and both can reach minimum $w(1)=2$. Further, maximum inter-element spacing constraint (MISC) array has been proposed in~\cite{MISC} to further reduce $w(1)$ to $w(1)=1$.

\par
Although many SA geometries with good properties have been proposed, only few papers discuss principles of designing SAs with desired properties. SNA and ANA are both presented with a detailed design process. However, they use set operations. As a consequence, it leads to a complicated discussion of many cases that need to be analyzed~\cite{SNA1,SNA2,ANA}. The design of MISC array is somewhat intuitive with the position set given directly~\cite{MISC}. In~\cite{SA_design}, an SA design method through fractal geometries is proposed which provides a trade-off between uniform DOF (uDOF), mutual coupling, and robustness. However, the fractal geometry still relays on the existing SAs to act as a generator. Besides, the number of sensors is exponentially increasing with the array order which makes it complicated for the design process to be practically useful.

\par
Motivations of this paper are as follows. First, ULAs are special types of arrays, which can be easily parameterized by only 3 parameters such as the initial position, inter-element spacing, and number of sensors. Second, properties of ULAs are easier to investigate than SAs. Third, the well known nested and coprime arrays are in fact SAs that consist of two sub-ULAs and can be regarded as special cases of a more generic SA design principle. Finally, there still exists no systematic SA design method.

\par
The main objective of this paper is to formulate an SA design principle, which we call as ULA fitting. The basic idea of ULA fitting is to design SA using concatenation of a series of ULAs. For example, the known SAs, such as nested and coprime arrays, are in fact combinations of two sub-ULAs with different spacing. However, the case when there are more than two sub-ULAs within one SA structure has not been investigated to the best of our knowledge. Although, ANA and MISC arrays contain more than two sub-ULAs, they are still special cases.

\par
The problem of modeling an SA with arbitrary sub-ULAs and further analyzing such model is then of interest. In this respect, the polynomial model~\cite{effective_aperture, polynomial_factorization, optimum_array_processing, Vorob3} is convenient for the purposes of this paper. Note that in~\cite{effective_aperture} and~\cite {polynomial_factorization}, the authors mainly focus on designing effective aperture equivalent to a ULA using sparse receive and transmit arrays. In \cite{effective_aperture}, based on the fact that the radiation pattern is the Fourier transform of the aperture function, the convolution of receive and transmit aperture functions is translated into the multiplication of their corresponding one-way radiation patterns. The work~\cite{polynomial_factorization} further extended the method of~\cite{effective_aperture} by utilizing polynomial factorization to design sparse periodic linear transmit and receive arrays. In this paper, 
the polynomial model is employed as a powerful tool which links the physical sensor array, DCA, and weight function. The polynomial model is established to analyze the case when an SA consists of arbitrary sub-ULAs. Based on investigation of the polynomial, the ULA fitting principle is formulated. Moreover, the main components of ULA fitting, new SA division, design criteria, and pseudo-polynomial equation are introduced. The ULA fitting enables to design SAs with low mutual coupling, closed-form expressions and large uDOF using pseudo-polynomials.

The main contributions are the following. (i)~We introduce a polynomial model to analyze DCA and establish basic relationship between physical sensor positions and the corresponding DCA and weight function. (ii)~We use the polynomial to model SA with arbitrary sub-ULAs and further analyze the constitute of corresponding DCA. Properties of each component in DCA are investigated, and based on these properties a novel SA design principle (ULA fitting) is proposed. (iii)~We develop a pseudo polynomial function for designing SAs which enables finding specific solutions for SA geometries with closed-form expressions. (iv) Two novel geometries with reduced mutual coupling are also proposed.

\par
The paper is organized as follows. Section~\ref{Preliminaries} explains basic models, DCA concept, and coupling leakage effect. Section~\ref{UF_and_PM} presents the idea of ULA fitting and establishes the polynomial model of SAs. In Section~\ref{UF_ANA}, the polynomial model is utilized to investigate the DCA of SAs. Section~\ref{UF_principle} gives the basic components of ULA fitting, criteria for parameter selection, and lower bound on uDOF. The design procedure along with two specific examples of the use of ULA fitting are given in Section~\ref{UF_example}. Numerical examples are presented in Section~\ref{NE}. Section~\ref{conclusion} concludes the paper.

\par
\section{Preliminaries}\label{Preliminaries}
\subsection{Signal Model}
Let us consider an array with $N$ sensors located at the positions $\mathbb{S}\times\lambda/2$, where $\lambda$ is the signal wavelength. Position set $\mathbb{S}$, referred to as the {\it normalized position set}, possesses unit underlying grid and is an integer set:
\begin{equation}
\mathbb{S} = \left\{ {{p_l}, \, l = 0,1, \cdots N-1} \right\}.
\label{position_set}
\end{equation}
For the array, the steering vector for a given direction $\theta$ is given as
$\mathbf{a}(\theta ) = [ {{e^{j(2\pi /\lambda ){p_0}\sin \theta }}}, {{e^{j(2\pi /\lambda ){p_1}\sin \theta }}}, $ $ \cdots {{e^{j(2\pi /\lambda ){p_{N - 1}}\sin \theta }}} ]^T$, where $ [\cdot ]^T$ is the transpose.

\par
Let us assume that $Q$ uncorrelated narrowband source signals are impinging on the array from azimuth directions ${\boldsymbol \theta} = [{\theta_1}, {\theta_2}, \ldots, {\theta_Q}]^T$, and $L$ snapshots are available. More specifically, there are source signals $\left\{s_q(l), \, q =1,2,\cdots,Q\right\}$ with powers $\left\{\sigma ^2_q, \, q =1,2,\cdots,Q \right\}$ where $\left\{l, \, l = 1,2, \cdots, L\right \}$. Then the received signal in snapshot  $l$ can be expressed as
\begin{equation}
{\mathbf{x}_l} = {\mathbf{A}}{\mathbf{s}_l} + {\mathbf{n}_l},
\label{received_data_one_snapshot}
\end{equation}
where $\mathbf{A}$ is the steering matrix with columns $\left\{ \mathbf{a}(\theta_i), \, i =1, \cdots,Q\right\}$, $\mathbf{s}_l= [s_1(l),\ldots,s_Q(l)]^{\rm{T}}$, and $\mathbf{n}_l$ is the white noise vector which is assumed to be independent from the sources.

\subsection{Difference Coarray}
The main building block in this paper is DCA. The essence of DCA is the utilization of the second order statistics, namely, the covariance matrix of the received signal. Using~(\ref{received_data_one_snapshot}), the covariance matrix of the received signal $\mathbf{x}_l$ is computed as
\begin{equation}
{\mathbf{R}_{\mathbf{xx}}}(l) = E[\mathbf{x}_l{\mathbf{x}_l^H}] = \mathbf{A}\mathbf{R}_{\mathbf{ss}}(l){\mathbf{A}^H} + \sigma_n^2 \mathbf{I},
\label{covariance_matrix}
\end{equation}
where $\mathbf{R}_{\mathbf{ss}}(l)=\rm{diag} \left[\sigma ^2_1,\sigma ^2_2,\cdots,\sigma ^2_Q\right]$ denotes the source covariance matrix, $\sigma ^2_n$ is the noise power, $(\cdot)^H$ is the Hermitian transpose, and $\rm{diag}[\cdot]$ forms a diagonal matrix from a vector. Vectorized version of~(\ref{covariance_matrix}) is expressed as
\begin{equation}
{\mathbf{f}}={\rm{vec}}({\mathbf{R}_{\mathbf{xx}}}) = ({\mathbf{A}}^*\odot{\mathbf{A}}){\mathbf{g}}+ \sigma_n^2 {\mathbf{1}}_n,
\label{vectorized_covariance_matrix}
\end{equation}
where ${\mathbf{g}} = \left[\sigma ^2_1,\sigma ^2_2,\cdots,\sigma ^2_Q\right]^T$, ${\mathbf{1}}_n={\rm{vec}}({\mathbf{I}}_N)$, $\odot$ is the Khatri-Rao product, and ${\rm{vec}}(\cdot)$ is the vectorization operator.

Comparing~(\ref{received_data_one_snapshot}) and~(\ref{vectorized_covariance_matrix}), ${\mathbf{f}}$ can be regarded as a received signal of a virtual sensor array whose manifold is expressed as $({\mathbf{A}}^*\odot{\mathbf{A}})$. This virtual sensor array is the well known DCA whose sensor positions are given by
\begin{equation}
\mathbb{D}= \left\{ {{p_m-p_n}, \, m,n = 0,1, \cdots N-1} \right\}.
\end{equation}

\begin{definition}
{\it (DOF)} Given an SA $\mathbb{S}$, the DOF is the cardinality of its DCA $\mathbb{D}$.
\end{definition}

\begin{definition}
{\it (uDOF)} Given an SA $\mathbb{S}$, the uDOF is the cardinality of the central ULA of its DCA $\mathbb{D}$, denoted as $\mathbb{U}$.
\end{definition}

\par
We discuss DCA here in two aspects, namely the coarray structure and the weight function, which are defined as follows.

\begin{definition}
{\it (Coarray Structure)} The coarray of an array $\mathbb{S}$ is the array geometry with sensor position set $\mathbb{D}$.
\end{definition}

\begin{definition}
{\it (Weight Function)} The weight function $w(n)$ of an array $\mathbb{S}$ is the number of sensor pairs with interval $n$.
\end{definition}

\par
For a given SA $\mathbb{S}$, its weight function $w(n)$ can be computed as~\cite{NESTED}
\begin{equation}
w(n)=b(n)\oplus b(-n),
\label{convolution_weight_function}
\end{equation}
where $b(n)$ is the binary expression of $\mathbb{S}$ given by
\begin{equation}
 b(n)={\left\{ {\begin{array}{*{20}{l}}
1,&n \in \mathbb{S}\\
0,&{\rm {elsewhere}}.\\
\end{array}} \right.}
\label{binary_expression_of_SA}
\end{equation}
It can be concluded from~(\ref{convolution_weight_function}) that the length of $w(n)$ is $2{\rm{max}}\{\mathbb{S}\}+1$.

\par
Using DCA, the DOF of an SA can be significantly increased. However, it is the uDOF that determins the ability of identifying uncorrelated sources~\cite{NESTED,ANA,COPRIME}. Therefore, we mainly focus on designing SAs with high uDOF and low mutual coupling.

\subsection{Coupling Leakage}
In practice, sensors with small separation space interfere with each other owing to energy radiation and absorption. It is known as mutual coupling. Therefore, a coupling matrix  $\mathbf{C}$ should be incorporated into~(\ref{received_data_one_snapshot}), that is,
\begin{equation}
{\mathbf{x}_l} = {\mathbf{C}} {\mathbf{A}}{\mathbf{s}_l} + {\mathbf{n}_l}.
\label{received_data_one_snapshot_with_coupling}
\end{equation}
Typically, mutual coupling is a result of many factors, e.g., humidity, operating frequency, adjacent objects, and so on, which leads to a complicated expression for $\mathbf{C}$~\cite{coupling1}. In this paper, only linear array is considered and, thus, a simplified $\mathbf{C}$ can be utilized. Based on the inverse proportion relationship between sensor interspace and coupling coefficients, an approximate $\mathbf{C}$ can be formulated using a B-banded mode as~\cite{SNA1,coupling1,coupling2,coupling3}
\begin{equation}
{\mathbf{C}}_{i,j}=\left\{{\begin{array}{*{20}{l}}
c_{|p_i-p_j|},&|p_i-p_j|\le B,\\
0,&{\rm{elsewhere}},
\end{array}}  \right.
\label{c_approximate}
\end{equation}
where $p_i, \, p_j\in \mathbb{S}$ and $c_a,a\in [0,B]$ are coupling coefficients, which satisfy the following relationships
\begin{equation}
\left\{{\begin{array}{*{20}{l}}
c_0=1\textgreater |c_1|\textgreater |c_2|\textgreater\cdots\textgreater|c_B|,\\
|c_g/c_h|=h/g.
\end{array}} \right.
\label{c_coefficients}
\end{equation}

\par
To evaluate the mutual coupling effect, the coupling leakage $L$ is introduced as
\begin{equation}
L=\frac {||{\mathbf{C}}-{\rm{diag}}({\mathbf{C}})||_F} {||{\mathbf{C}}||_F},
\label{coupling_leakage}
\end{equation}
where $\| \cdot \|_F$ stands for the Frobenius norm of a matrix. Qualitatively, the higher $L$ is, the heavier mutual coupling is. Considering the definition of weight function (Definition~4) together with~(\ref{c_approximate}),~(\ref{c_coefficients}), and~(\ref{coupling_leakage}), it can be concluded that the mutual coupling is mainly dominated by the weight function for small coarray indexes, i.e., $w(1)$, \, $w(2)$, and $w(3)$ \cite{SNA1, SNA2}. It is simply because the non-zero weights for small indexes in $\mathbb{D}$ indicate small separation space between sensors in $\mathbb{S}$ which are interfering with each other.

\par
\section{ULA Fitting: Idea and Polynomial Model}\label{UF_and_PM}
\subsection{General Idea of ULA Fitting}
The general idea of ULA fitting is to design SAs from sub-ULAs. Indeed, the well known SA structures, e.g., coprime and nested arrays can be regarded as a combination of two sub-ULAs~\cite{NESTED,COPRIME}. However, the case when an SA is composed of more sub-ULAs still remains unclear. Here we utilize and further develop the polynomial model for an array as an analitic tool for ULA fitting analysis.
\begin{figure}[htp]
\centering
\centerline{\includegraphics[width=1.0\columnwidth,height=1.8cm]{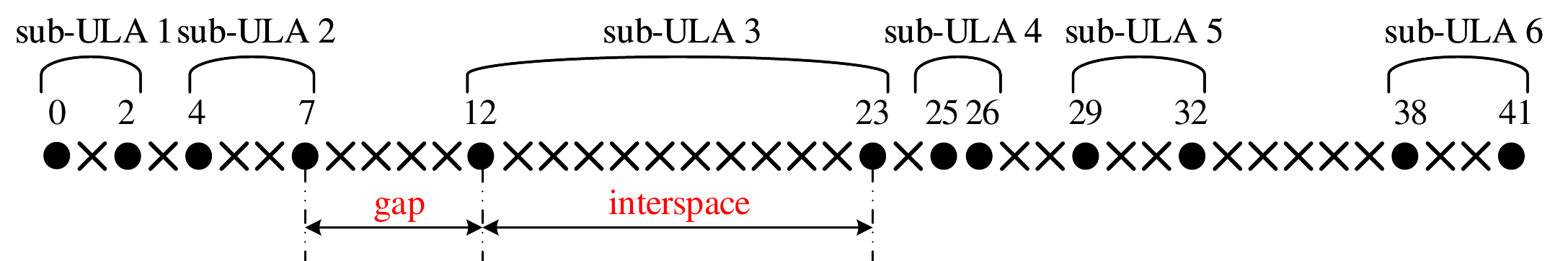}}
\centering
\caption{ULA fitting using 6 sub-ULAs.}
\label{illustration_n2}
\end{figure}

\par
To begin with, we shall introduce several concepts. Fig.~\ref{illustration_n2} shows one specific SA geometry explained later in the paper. It consists of 6 sub-ULAs. There are two concepts that will be used throughout the paper. The inter-element spacing within each sub-ULA is referred to as {\it interspace}, and the interval between adjacent sub-ULAs is referred to as {\it gap}. More specifically, a gap between two sub-ULAs is the distance between the last sensor in the former sub-ULA and the first sensor of the latter sub-ULA. The sequence of sub-ULAs is organized by the initial position of each sub-ULA. The interspace, number of sensors, aperture, and initial position of sub-ULA $i$ are denoted by $S_i,N_i,AP_i$, and $I_i$, respectively. For example, $S_1=5$, $N_2=3$, $AP_3=10$, and $I_4=12$ indicate that the interspace of sub-ULA $1$ is $5$, number of sensors in sub-ULA 2 is 3, aperture of sub-ULA 3 is 10, and initial position of sub-ULA 4 is 12. Gap between sub-ULA $i$ and sub-ULA $j$ is expressed as ${\rm{gap}}_{i,j}$. For example, the gap between sub-ULA $1$ and sub-ULA $2$ is $\rm{gap}_{1,2}$. We just consider the gaps between adjacent sub-ULAs. Other gaps can be then computed using adjacent gaps and apertures. Within one SA, the following relationships can be easily established
\begin{equation}
\left\{{\begin{array}{*{20}{l}}
AP_j = S_j \times (N_j-1),\\
I_1=0,\\
I_j = I_{j-1} + AP_{j-1} + \rm{gap}_{j-1,j}, \, j\ge2,\\
{\rm{gap}}_{m,n}=\sum _{i=m}^{n-1}{\rm{gap}}_{i,i+1}+\sum _{j=m+1}^{n-1}AP_{j}, \, m\textless n.
\end{array}} \right.
\label{ula_fitting_relationships}
\end{equation}
Based on~(\ref{ula_fitting_relationships}), a ULA can be determined by only $3$ parameters. Therefore, we represent sub-ULA $i$ as a triple $\left\{I_i, S_i, N_i \right\}$.

\par
\subsection{Polynomial Model}
Based on $z$-transform and discrete Fourier transform (DFT), polynomial model for an array has been utilized for designing desired radiation patterns of linear arrays~\cite{optimum_array_processing, effective_aperture, polynomial_factorization}. The DFT of $b(n)$ can be expressed as
\begin{equation}
B(k) = \sum _{n=0}^{{\rm{max}}\{\mathbb{S}\}}b(n){\rm{e}}^{\frac {-j2\pi kn} {{\rm{max}}\{\mathbb{S}\}+1}}, \, k=0,\cdots,{\rm{max}}\{\mathbb{S}\}.
\label{DFT_B}
\end{equation}
Substituting $x={\rm{e}}^{\frac {-j2\pi k} {{\rm{max}}\{\mathbb{S}\}+1}}$ in~(\ref{DFT_B}), the polynomial expression for SA $\mathbb{S}$ in~(\ref{DFT_B}) is given as
\begin{equation}
P_{\rm{SA}}(x) = \sum _{n=0}^{{\rm{max}}\{\mathbb{S}\}}b(n)x^n.
\label{polynomial_model_arbitrary_SA}
\end{equation}
Using~(\ref{convolution_weight_function}) and~(\ref{polynomial_model_arbitrary_SA}), the DCA corresponding to SA $\mathbb{S}$ can be then expressed in terms of the following polynomial model
\begin{equation}
\begin{aligned}
&P_{\rm{DCA}}(x) = P_{\rm{SA}}(x)\times P_{\rm{SA}}(x^{-1})\\
&=\sum _{n=-{\rm{max}}\{\mathbb{S}\}}^{{\rm{max}}\{\mathbb{S}\}}w(n)x^n.
\label{polynomial_model_DCA}
\end{aligned}
\end{equation}
Equation~(\ref{polynomial_model_DCA}) establishes an important relationship between the original SA $\mathbb{S}$, corresponding DCA $\mathbb{D}$, and weight function $w(n)$. In $P_{\rm{DCA}}(x)$, there is a one-to-one correspondence between each exponent and coarray index $\mathbb{D}$, while coefficients correspond to the weight function $w(n)$. Note that~(\ref{polynomial_model_DCA}) in fact contains all the information about the DCA, namely, coarray structure and weight function. However, as we have analyzed, to design an SA with low mutual coupling, it is important to focus only on $w(1)$, $w(2)$, and $w(3)$. Therefore, we discuss the coarray structure and the weight function separately. Hence, two simplified versions of the traditional operators $\times$ and $+$ are utilized for convenience. The simplified operators, $\widetilde\times$ and $\widetilde+$ inherit all the computation rules of $\times$ and $+$, but they omit coefficients. For example, $(x+x^2)\times(x+x^2)=x^2+2x^3+x^4$ while $(x+x^2)\widetilde\times(x+x^2)=x^2+x^3+x^4$. Besides, $(x+x^2)+(x^2+x^3)=x+2x^2+x^3$ while $(x+x^2)\widetilde+(x^2+x^3)=x+x^2+x^3$.

\par
Consider sub-ULA $1$ with parameters $\left\{ I_1,S_1,N_1\right\}$ that can be expressed using polynomial model as
\begin{equation}
\begin{aligned}
P_{\rm{sub1}}(x) &=  x^{I_1} + x^{I_1+S_1} +  \cdots +x^{I_1+S_1(N_1-1)}\\
&= \left( x^0 + x^{S_1} +  \cdots +x^{S_1(N_1-1)} \right){x^{I_1}}.
\end{aligned}
\label{arbitrary_ULA_A}
\end{equation}
Equation~(\ref{arbitrary_ULA_A}) simply demonstrates the fact that any ULA can be regarded as a shifted version of a special type of ULAs whose initial position is $0$. Herein, we give these type of ULAs new definition and notation.
\begin{definition}
{\it (Prototype Array)} Prototype array is a ULA with first sensor located at reference point.
\end{definition}

\par
Particularly, we present a simplified notation for prototype arrays which only contains interspace and number of sensors. Considering the prototype array of the aforementioned sub-ULA $1$ which has $N_1$ sensors and interspace $S_1$, we have the following notation
\begin{equation}
P_{\rm{proto}}^{\rm{A}}\left\{S_1, N_1 \right\} = x^0 + x^{S_1} +  \cdots +x^{S_1(N_1-1)}.
\label{proto_ULA_A}
\end{equation}
\par

\subsection{Properties of Polynomial Model}

\subsubsection{Shifting Property} \label{shift}
Consider ULA $i$ with parameters $\left\{ I_i, S_i, N_i \right\}$ and the corresponding polynomial expression $P_{i} (x)$, and also ULA $j$ with polynomial expression $P_{j} (x) = P_{i} (x) \times x^{n}$. Then ULA $j$ is the $n$-shifted version of ULA $i$ with parameters $\left\{ I_i+n, S_i, N_i \right\}$, where $n$ can be any integer. The shifting property in fact clarifies the mapping relationship between polynomial multiplication and array shifting.

\subsubsection{Duality Property}
For arbitrary SA, there always exists a dual array which shares the same DCA structure with it. Assuming an SA with arbitrary configuration and its corresponding polynomial expression $P_{\rm{SA}}(x)$, the method for obtaining its dual array is
\begin{equation}
\begin{aligned}
P_{{\rm{SA}}}^{\rm{dual}}(x) = P_{\rm{SA}}(x^{-1})\times x^{{\rm{max}}\left\lceil {P_{\rm{SA}}(x)} \right\rceil},
\label{polynomial_duality}
\end{aligned}
\end{equation}
where ${{\rm{max}}\left\lceil {P_{\rm{SA}}(x)} \right\rceil}$ represents the maximum exponent value of $P_{\rm{SA}}(x)$.

\par
The proof of this property is straightforward. The DCA for this SA can be expressed as
\begin{equation}
{P_{\rm{DCA}}^{\rm{SA}}(x)} = {P_{\rm{SA}}(x)}\times{P_{\rm{SA}}(x^{-1})} ,
\label{DC_of_array}
\end{equation}
while for the dual array, using~(\ref{polynomial_duality}){\color{red},} the DCA can be computed as
\begin{equation}
\begin{aligned}
{P_{\rm{DCA}}^{\rm{dual}}(x)} &= {P^{\rm{dual}}_{\rm{SA}}(x)}\times{P^{\rm{dual}}_{\rm{SA}}(x^{-1})}
\\&=\left\{ { P_{\rm{SA}}(x^{-1})\times x^{{\rm{max}}\left\lceil {P_{\rm{SA}}(x)} \right\rceil }} \right\}
\\&\times \left\{ { P_{\rm{SA}}(x)\times x^{{\rm{max}}\left\lceil {P_{\rm{SA}}(x^{-1})} \right\rceil}} \right\}
\\&={P_{\rm{SA}}(x)}\times{P_{\rm{SA}}(x^{-1})} = {P_{\rm{DCA}}^{\rm{SA}}(x)}.
\label{DCA_of_dual}
\end{aligned}
\end{equation}

\section{Analyzing ULA Fitting using Polynomial}\label{UF_ANA}
Now that we have established the polynomial model for an array, we use it for ULA fitting analysis. Suppose an SA is composed of $n$ sub-ULAs, then this SA can be modeled using polynomial as
\begin{equation}
P_{\rm{SA}}(x) = P_{\rm{sub1}}(x) + P_{\rm{sub2}}(x) + \cdots + P_{{\rm{sub}}(n)}(x),
\label{polynomial_of_sparse_array}
\end{equation}
thus yielding the corresponding DCA expression
\begin{equation}
\begin{aligned}
P_{\rm{DCA}}^{\rm{SA}}(x) &= P_{\rm{SA}}(x) \times P_{\rm{SA}}(x^{-1}) \\
&= \underbrace{\sum _{i=1}^{n}P_{{\rm{sub}}(i)}(x) \left\{ P_{\rm{SA}}(x^{-1})-P_{{\rm{sub}}(i)}(x^{-1})\right\}}_{\text{IDCAs}} \\
&+ \underbrace{\sum _{i=1}^{n}P_{{\rm{sub}}(i)}(x)P_{{\rm{sub}}(i)}(x^{-1})}_{\text{SDCAs}}.
\label{polynomial_of_sparse_FDCA}
\end{aligned}
\end{equation}
Without loss of generality, as shown in~(\ref{polynomial_of_sparse_FDCA}), a DCA can be divided into two parts, namely self-difference co-arrays (SDCAs) and inter-difference co-arrays (IDCAs). Besides, if an SA is composed of $n$ sub-ULAs, the corresponding DCA will have $n$ SDCAs and ${\rm{A}}_n^2$ IDCAs.

\par
Further, we investigate SDCA structures, IDCA structures, and weight function successively by considering two sub-ULAs, namely, sub-ULA 1 and sub-ULA 2 with parameters $\left\{ I_1, S_1, N_1 \right\}$, $\left\{ I_2, S_2, N_2 \right\}$, respectively, and $\rm{gap}_{1,2}$. Note that for structure analysis, we use the simplified operators for convenience.

\subsection{Analysis of SDCA Structures}
Using polynomial model, we first investigate the SDCA structures of sub-ULA 1 and sub-ULA 2, which yields
\begin{equation}
\begin{aligned}
&{P_{\rm{SDCA}}^{\rm{sub} 1}}(x) = {P_{\rm{sub 1}}(x)} \widetilde\times {P_{\rm{sub 1}}(x^{-1})}
\\ = & \left( {{x^0} + \cdots + x^{(N_1-1)S_1}} \right)\widetilde\times\left( {{x^0} + \cdots+ x^{-(N_1-1)S_1}}\right)
\\ = & x^{-(N_1-1)S_1} + \cdots + x^0 + \cdots + x^{(N_1-1)S_1},
\label{ULA_1_SDCA}
\end{aligned}
\end{equation}
and
\begin{equation}
\begin{aligned}
&{P_{\rm{SDCA}}^{\rm{sub} 2}}(x) = {P_{\rm{sub 2}}(x)} \widetilde\times {P_{\rm{sub 2}}(x^{-1})}
\\ = & \left( {{x^0} + \cdots + x^{(N_2-1)S_2}} \right)\widetilde\times\left( {{x^0} + \cdots+ x^{-(N_2-1)S_2}}\right)
\\ = & x^{-(N_2-1)S_2} + \cdots + x^0 + \cdots + x^{(N_2-1)S_2}.
\label{ULA_2_SDCA}
\end{aligned}
\end{equation}

\par
Expressions (\ref{ULA_1_SDCA}) and (\ref{ULA_2_SDCA}) imply several properties for SDCAs, which can be summarized as follows. 
(i)~The SDCA of an arbitrary linear array geometry is symmetric.
(ii)~The SDCA of an $N$-element ULA contains $2N-1$ distinct sensors, while the positive set is same as its prototype array.
(iii)~The initial position of any ULA has no effect on its SDCA.
(iv)~The period of an SDCA for any ULA is the same as its interspace.
In addition to these properties of SDCAs, in designing process, we can just consider the prototype arrays of each sub-ULA for convenience.

\par
\subsection{Analysis of IDCA Structures}
There are two IDCA structures between sub-ULA 1 and sub-ULA 2, namely IDCA $12$ and IDCA $21$ with the following expressions
\begin{equation}
\left\{{\begin{array}{*{20}{l}}
P_{\rm{IDCA12}}(x) &= {P_{\rm{sub 2}}(x)} \widetilde\times {P_{\rm{sub 1}}(x^{-1})},\\
P_{\rm{IDCA21}}(x) &= {P_{\rm{ULA 1}}(x)} \widetilde\times {P_{\rm{ULA 2}}(x^{-1})}
\\ &= P_{\rm{IDCA12}}(x^{-1}).
\end{array}} \right.
\label{Definition_IDCA}
\end{equation}
Clearly, IDCA $12$ and IDCA $21$ are symmetric to each other. Further computing IDCA $12$, yields
\begin{equation}
\begin{aligned}
&P_{\rm{IDCA12}}(x) = {P_{\rm{sub 2}}(x)} \widetilde\times {P_{\rm{sub 1}}(x^{-1})}
\\ = & \underbrace{ P_{\rm{proto}}^{\rm{sub1}}\left\{S_1,N_1\right\}}_{\text{prototype of sub-ULA 1}}
 \widetilde\times\underbrace{ \left( {{x^0} + \cdots+ x^{(N_2-1)S_2}}\right)}_{\text{$N_2$ terms, period $S_2$}}x^{\rm{gap}_{12}}.
\label{ULA_IDCA12}
\end{aligned}
\end{equation}

Based on~(\ref{ULA_IDCA12}), it can be seen that the initial position of IDCA $12$ is determined by $\rm{gap}_{12}$. Besides, based on the aforementioned shifting property, (\ref{ULA_IDCA12}) provides us with a new insight about the IDCAs between two sub-ULAs which in fact implies a transfer process. One sub-ULA provides the prototype array to be transferred, while the other sub-ULA determines transfer times and transfer period. By default, the sub-ULA with larger interspace provides the transfer times and period.

\par
Without loss of generality, let us set $S_2 \textgreater S_1$, i.e., sub-ULA~2 provides the transfer times and period. Then, based on~(\ref{ULA_IDCA12}), if $S_2 \textgreater (N_1-1)S_1$, i.e., the interspace of sub-ULA~2 is larger than the aperture of sub-ULA 1, the number of sensors within each period of IDCA~$12$ is $N_1$. This fact is summarized then as the following proposition.

\begin{proposition}
If the interspace of the transfer sub-ULA is larger than the aperture of the prototype sub-ULA, then each period of the corresponding IDCAs possesses the same number of sensors with the prototype sub-ULA.
\label{proposition_coarray_number}
\end{proposition}

\par
Additionally, we have established that the gap between two sub-ULAs can determine the initial position of their IDCAs. Thus, by properly designing the gap between two sub-ULAs, we can perfectly acquire two IDCAs that are symmetrically distributed on both sides of the reference point. Thus, we can just choose the IDCA that is located at the positive side for convenience of the analysis.

\par
Based on the analysis above, we summarize the properties of IDCAs as follows.
(i)~Two IDCAs between two sub-ULAs are symmetric to the reference point.
(ii)~The initial position of IDCAs are determined by the gap between the corresponding sub-ULAs.
(iii)~The IDCAs between two sub-ULAs can be regarded as a transferring process, while one sub-ULA provids the prototype array to be transferred, and the other sub-ULA determines transfer times and period.

\par
\subsection{Example}
\begin{figure}[htp]
\centering
\centerline{\includegraphics[width=1\columnwidth,height=5cm]{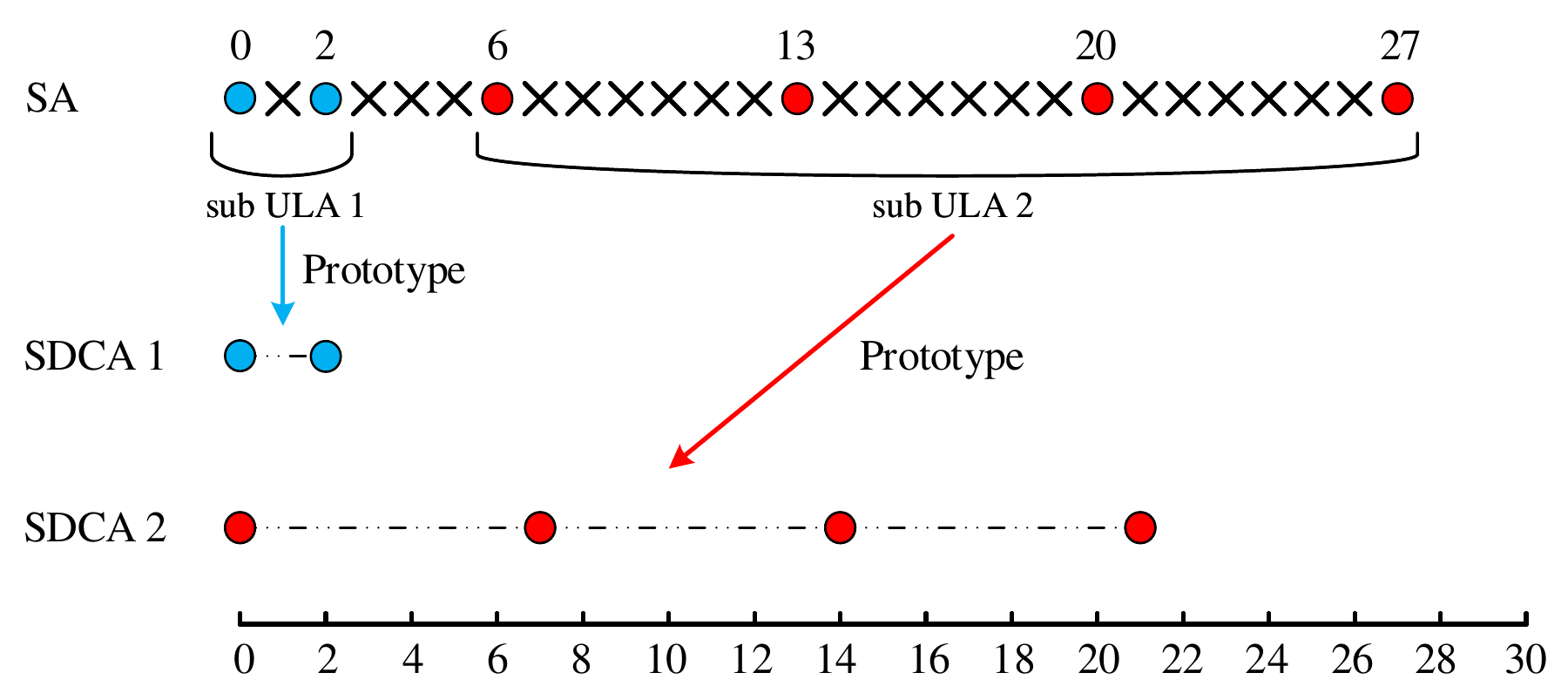}}
\centering
\caption{Relationship between SA and the corresponding SDCA structures. ULAs in the same color indicate that they have the same prototype array.}
\label{SA_VS_SDCA}
\end{figure}

\begin{figure}[htp]
\centering
\centerline{\includegraphics[width=1\columnwidth,height=5cm]{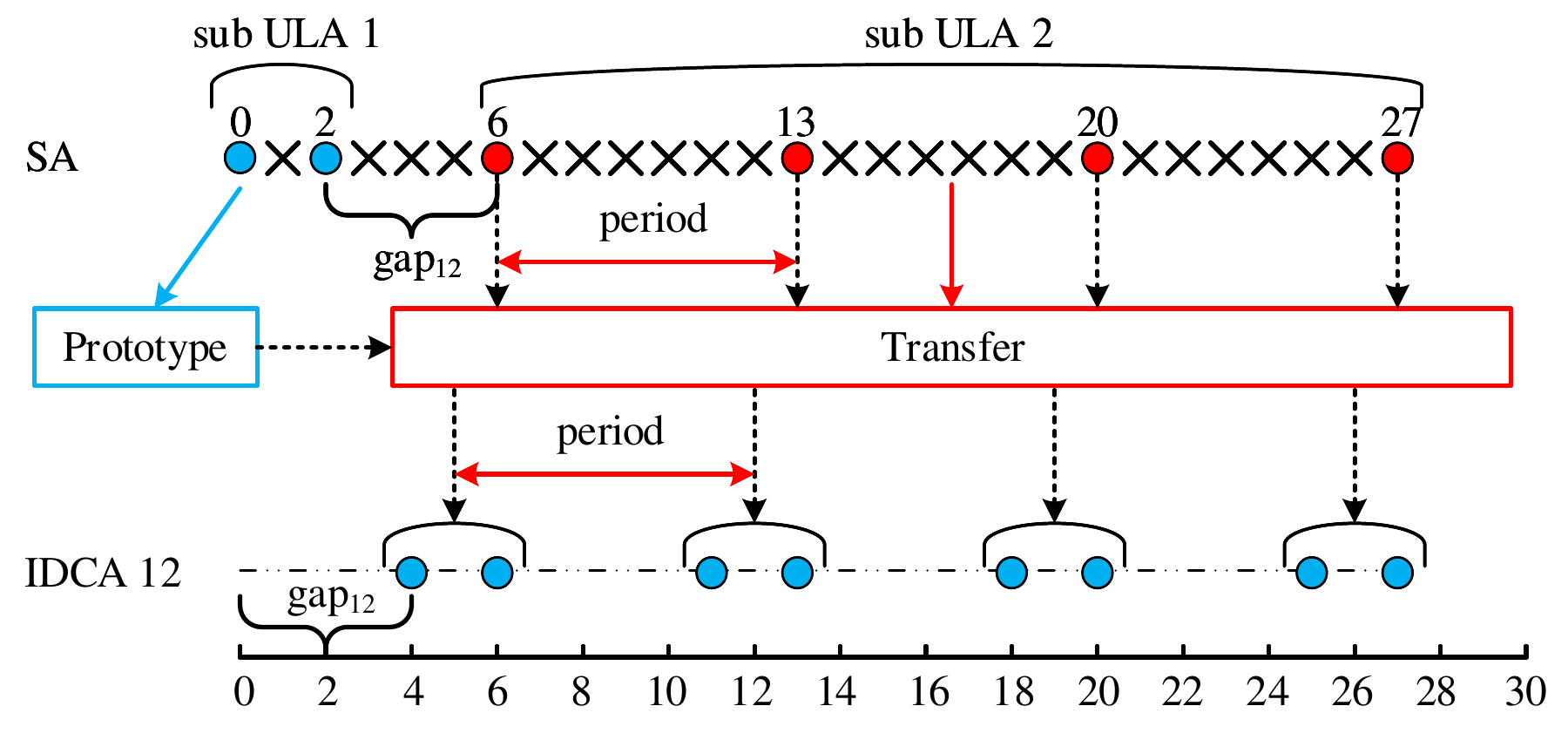}}
\centering
\caption{Relationship between SA and IDCA~$12$. Sub-ULA~$1$ is the prototype sub-ULA and sub-ULA~$2$ is the transfer sub-ULA, $\rm{gap}_{1,2}$ determines the initial position of IDCA $12$. ULAs in the same color indicate that they have the same prototype array.}
\label{SA_VS_IDCA}
\end{figure}

To better comprehend the properties of SDCA's and IDCA's structures, an example of a specific SA structure is presented. The SA consists of 2 sub-ULAs with parameters $\left\{ 0,2,2\right\}$, $\left\{6,7,4\right\}$ and $\rm{gap}_{1,2}=4$. The SA structure versus the SDCAs and the SA versus IDCA $12$ are illustrated in~Figs.~\ref{SA_VS_SDCA} and~\ref{SA_VS_IDCA}, respectively.

\par
\subsection{Analysis of Weight Function}
As argued before, only $w(1)$, $w(2)$, and $w(3)$ are of significance. 
For other coarray indexes, it is more important to investigate their existence rather than specific weights for constructing a long consecutive coarray structures.

\begin{proposition}
The SDCA of a ULA does not introduce weight function for coarray indices smaller than its interspace.
\label{proposition_SDCA_wf}
\end{proposition}

{\it Proof:}
Consider ULA $i$ with parameters $\left\{I_i, S_i, N_i \right\}$ and corresponding polynomial expression $P_{i}(x)$. The weight function contributions given by its SDCA are the coefficients of $P^{i}_{\rm{SDCA}}(x)$, which satisfy
\begin{equation} 
w(n) {\rm{ = }}
\left\{{\begin{array}{*{20}{l}}
N_{i} - m,&n=m\times S_{\rm{i}},m = 0,1,\cdots,N_{i},\\
0,&{{\mathrm{elsewhere.}}}
\end{array}} \right.
\label{proof_SDCA_wf}
\end{equation}
Therefore, a ULA can only introduce non-zero weights for coarray indexes that are multiples of its interspace and no weights for smaller indexes.

\begin{proposition}
The IDCA between two sub-ULAs does not introduce weights for coarray index smaller than their gap.
\label{proposition_IDCA_wf}
\end{proposition}
\par
Proposition~\ref{proposition_IDCA_wf} is straightforward because the initial position of IDCA is determined by the gap between the corresponding sub-ULAs.
\par
The ULA fitting principle can now be introduced based on Propositions~\ref{proposition_SDCA_wf} and~\ref{proposition_IDCA_wf} along with properties of SDCAs and IDCAs.

\section{ULA Fitting}\label{UF_principle}
\subsection{General Components of ULA Fitting}
The objective is to specify the ULA fitting principle based on which one can construct SAs with closed-form expressions, large uDOF, and low mutual coupling. In ULA fitting, an SA generally consists of base layer, transfer layer and addition layer. First, we introduce these layers.

\par
\subsubsection{The Base Layer}
\begin{definition}
{\it (The Base Layer)} The base layer consists of sub-ULA(s) with the same structure that pad each period (provided by the transfer layer) to realize a dense coarray.
\end{definition}

\par
Number of sub-ULAs in base layer is $N_{\rm{base}}$, and each sub-ULA has parameters $\left\{N_{\rm{b}}, S_{\rm{b}}, AP_{\rm{b}}\right\}$. The following properties hold for the base layer.
(i)~$S_{\rm{b}}$ is fixed and $N_{\rm{b}}$ is increasing.
(ii)~The base layer can consist of more than one sub-ULA, each sub-ULA has the same prototype array.
(iii)~The base layer has the ability of constructing a dense ULA.
(iv)~Aperture of base layer is equivalent to $AP_{\rm{b}}$.

\par
\begin{figure}[htp]
\centering
\centerline{\includegraphics[width=1\columnwidth,height=4.5cm]{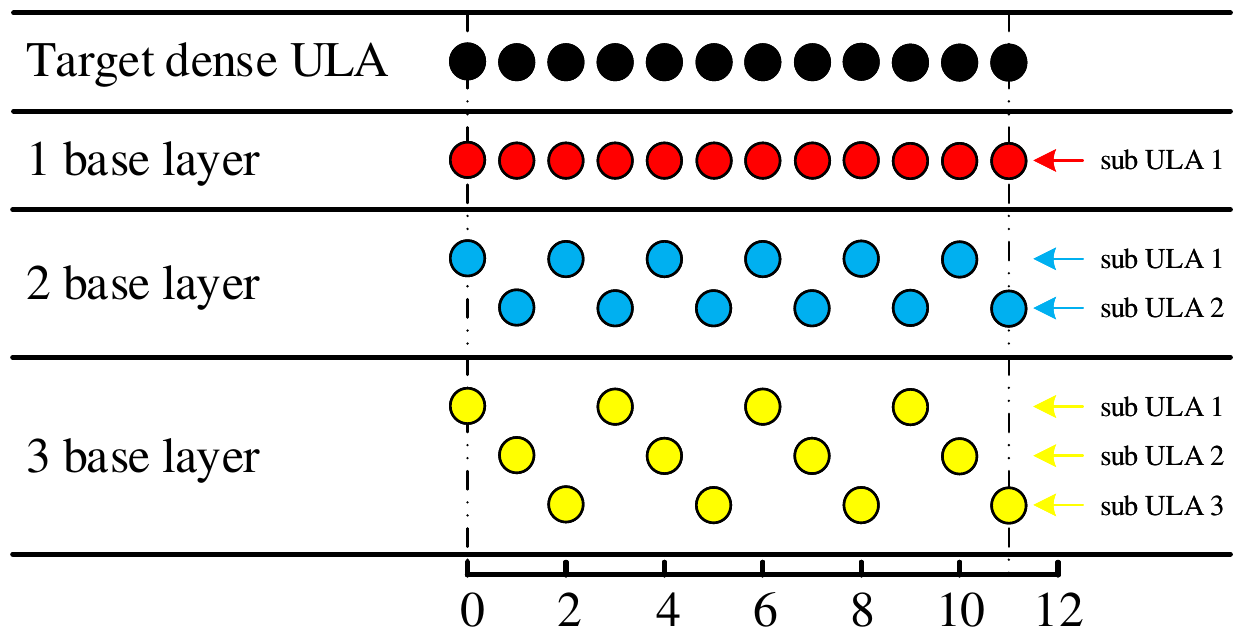}}
\centering
\caption{Illustration of utilizing $1$, $2$ and $3$ base layers for realizing the target dense coarray space.}
\label{base_layer}
\end{figure}

The base layer is named by the interspace of its sub-ULAs. For example, sub-ULAs within one base layer have interspace~$1$. In ULA fitting, the main function of the base layer is to cover specific coarray space. Fig.~\ref{base_layer} shows how to utilize $1$, $2$, and $3$ base layers to cover a target dense ULA. Generally speaking, as Fig.~\ref{base_layer} indicates, the base layer with larger interspace always has more sub-ULAs. Besides, the arrangement of the sub-ULAs pertaining to one base layer is realized by properly designed gaps.
\par
The great significance of developing base layer with large interspace is to reduce the mutual coupling. This is quite straightforward due to the augmented interspace and Proposition~\ref{proposition_SDCA_wf}. For instance, as illustrated in~Table~\ref{weight_function_summarize}, the mutual coupling is reduced successively with $1$, $2$, and $3$ base layers. Note that we just consider here the mutual coupling within the base layer. Influence of gaps is omitted.

\begin{table}[htp]
\centering
\caption{Weight contributions of the three base layer cases in~Fig.~\ref{base_layer} }
\renewcommand{\arraystretch}{0.8}
\label{weight_function_summarize}
    \begin{tabular}{lccc}
    \toprule
    layer                                           & w(1)      & w(2)     &w(3)     \\
    \midrule
    $1$ base layer                                  &11         & 10       &9        \\
    \midrule
    $2$ base layer                                  &0          & 10       &0        \\
    \midrule
    $3$ base layer                                  &0           & 0       &9        \\
    \bottomrule
    \end{tabular}
\end{table}

\subsubsection{The Addition Layer}
\hfill
\par
Propositions~\ref{proposition_SDCA_wf} and~\ref{proposition_IDCA_wf} in fact explain the purpose of having the addition layer. For example, if $S_{\rm{b}} =3$ is selected, it is of importance to complement weights $w(1)$ and $w(2)$ in order to get large uDOF. One possible approach is to set several gaps as $1$ and $2$. However, this approach leads to a stiff design process. Another method is to introduce sub-ULAs (which are in fact the addition layers) equiped only with two sensors and interspaces $1$ and $2$. We control the values of $w(1)$, $w(2)$, and $w(3)$ mainly by the second method. However, it should be pointed out that the purpose of the addition layer is not confined to complement the weight function.

\begin{definition}
{\it (The Addition Layer)} The addition layer consists of sub-ULA(s) that pad each period (provided by the transfer layer) together with the base layer to realize a tense coarray.
\end{definition}

\par
Number of sub-ULAs in addition layer is $N_{\rm{addition}}$. Each sub-ULA in addition layer is parameterized as $\left\{N_{\rm{a}}(i),S_{\rm{a}}(i),AP_{\rm{a}}(i)\right\}$ where $i=1,\dots,N_{\rm{addition}}$. The following properties must hold for the addition layer.
(i)~The addition layer is not indispensable.
(ii)~The addition layer can complement weight function in particular coarray indexes.
(iii)~The addition layer can complement some holes within each period.
(iv)~$N_{\rm{a}}(i), \, \forall i$ are fixed (in most cases $N_{\rm{a}}(i)=2$), while there is no particular requirement for selecting $S_{\rm{a}}(i)$ (in most cases $S_{\rm{a}}(i)$ are fixed as well).

\subsubsection{The Transfer Layer}
\begin{definition}
{\it (The Transfer Layer)} The transfer layer is the sub-ULA that provides the transfer times and period.
\end{definition}

\par
Parameters of the sub-ULA in transfer layer are denoted by $\left\{N_{\rm{t}}, S_{\rm{t}}, AP_{\rm{t}} \right\}$. The following properties must hold at the transfer layer.
(i)~$S_{\rm{t}}$ is determined by $N_{\rm{b}}$ and $N_{\rm{a}}$, and it is larger than $AP_{\rm{a}}$ and $AP_{\rm{b}}$, while $N_{\rm{t}}$ is increasing.
(ii)~An SA structure contains only one transfer layer.
(iii)~The transfer layer contains only one sub-ULA.
\par

Properties~(i)~and~(ii) follow from Proposition~\ref{proposition_coarray_number}. Property~(i) guarantees that $AP_{\rm{t}}$ dominates the whole aperture of the SA and further leads to a new SA domain division principle later shown in this paper.

\par
We have introduced now all the components for SA design by ULA fitting. Note that, in the SAs examples designed in this paper (see Section~\ref{UF_example}), we only consider the case of using one base layer. SA design with multiple base layers has special requirements for transfer layer, which we leave as a future work. Reconsidering~(\ref{polynomial_of_sparse_array}) and~(\ref{polynomial_of_sparse_FDCA}), suppose that an SA consists of $n$ sub-ULAs. There will be $n$ SDCAs and $C_n^2 = n! / 2! (n-2)! = (n-1) n / 2$ IDCAs in DCA domain (positive side). These coarrays in DCA domain are disorganized and hard to analyze. Hence, we provide new divisions in both SA and DCA domains for convenience of the analysis.

\begin{figure*}[htp]
\centering
\centerline{\includegraphics[width=1.8\columnwidth,height=7.7cm]{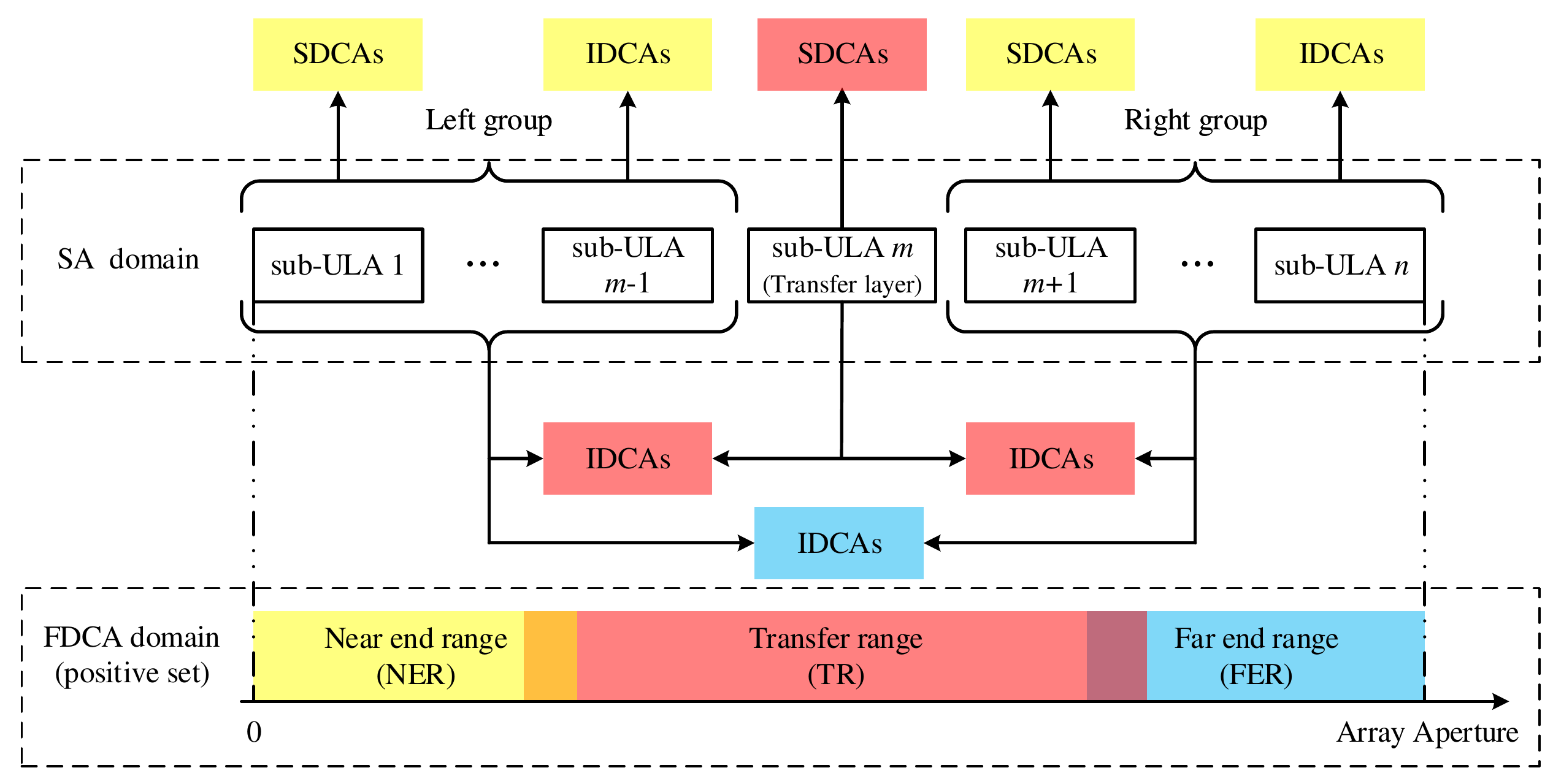}}
\caption{Illustration of the new array division in both SA domain and FDCA domain when ULA fitting is applied. Rectangles of the same color indicate the mapping relationships. Mixed colors indicate that the adjacent ranges are overlapped.}
\label{NET_TT_FET}
\end{figure*}

\par
\subsection{New Array Division and Parameter Selection}
\subsubsection{New SA and DCA Division}
\quad \par
Suppose that an SA is composed of $n$ sub-ULAs where the $m{\rm{th}}$ sub-ULA is selected as the transfer layer. Then the SA domain can be divided into three partitions, separated by the transfer layer, as illustrated in~Fig.~\ref{NET_TT_FET}, namely the left group (LG), the transfer layer (TL), and the right group (RG). The LG, TL, and RG can be expressed by polynomials, respectively, as
\begin{equation}
\left\{{\begin{array}{*{20}{l}}
P_{\rm{LG}}(x)={\sum _{i=1}^{m-1}P_{{\rm{sub}}_i}},\\
P_{\rm{TL}}(x)=P_{{\rm{sub}}m}(x),\\
P_{\rm{RG}}(x)={\sum _{i=m+1}^{n}P_{{\rm{sub}}_i}}.
\end{array}} \right.
\label{polynomial_LG_TL_RG}
\end{equation}

All the SDCAs and IDCAs of the SA can be easily mapped into three ranges in DCA domain, namely the near end range (NER), the transfer range (TR), and the far end range (FER). In~Fig.~\ref{NET_TT_FET}, blue rectangles indicate that the IDCAs in FER are generated by the sub-ULAs choosen separately from LG and RG. The remaining SDCAs and IDCAs can be further divided into two partitions. Obviously, based on the properties of SDCA and IDCA structures, all the coarrays related to the transfer layer share the same period $S_{\rm{t}}$. Hence, these $n$ coarrays, namely one SDCA and thus $n-1$ IDCAs, are assigned to the TR as illustrated by red rectangles in~Fig.~\ref{NET_TT_FET}. Other coarrays, shown as yellow rectangles in~Fig.~\ref{NET_TT_FET}, are assigned to the NER. It is easy to check that all the IDCAs in FER have initial positions larger than $AP_{\rm{t}}$, which is in fact the criteria of the new division. The polynomial model of NER, TR, and FER are, respectively, given as
\begin{equation}
\begin{aligned}
\left\{{\begin{array}{*{20}{l}}
P_{\rm{NER}}(x)&=(\sum _{i=1}^{n}P_{{\rm{SDCA}}_i}-P_{{\rm{SDCA}}_m})\\
&\quad \widetilde+(\sum _{i=1}^{m-2}\sum _{j=i+1}^{m-1}P_{{\rm{IDCA}}_{i,j}})\\
&\quad \widetilde+(\sum _{i=m+1}^{n-1}\sum _{j=i+1}^{n}P_{{\rm{IDCA}}_{i,j}}),\\
P_{\rm{TR}}(x)&=P_{{\rm{SDCA}}_m}\widetilde+(\sum _{i=1}^{m-1}P_{{\rm{IDCA}}_{i,m}})\\
&\quad \widetilde+(\sum _{j=m+1}^{n}P_{{\rm{IDCA}}_{m,j}}),\\
P_{\rm{FER}}(x)&=\widetilde\sum _{i=1}^{m-1}\widetilde\sum _{j=m+1}^{n}P_{{\rm{IDCA}}_{i,j}}.\\
\end{array}} \right.
\end{aligned}
\label{polynomial_NER_TR_FER}
\end{equation}

\par
By elaborately designing the interspaces and gaps within one SA, it is possible to guarantee a hole-free DCA. Using polynomial model, the constraint/requirement of a hole free DCA can be expressed as
\begin{equation}
P_{\rm{NER}}(x) \widetilde+ P_{\rm{TR}}(x) \widetilde+ P_{\rm{FER}}(x)={P_{\rm{proto}}\left\{1,{{\rm{max}}\left\lceil {P_{\rm{SA}}} \right\rceil} + 1 \right\}}.\\
\label{hole_free_FDCA}
\end{equation}

Note that~equation (\ref{hole_free_FDCA}) has multiple solutions. However, the need to guarantee the consecutiveness of the whole DCA may increase the coupling leakage. Thus, a relaxed criteria can be used for achieving a compromise between uDOF and coupling leakage. The relaxed criteria then should guarantee that the consecutive range contains $AP_{\rm{t}}$. It is based on the fact that $AP_{\rm{t}}$ dominates the aperture of the entire SA. Therefore, the relaxed criteria together with corresponding constraints can be expressed using polynomial model as, for example
\begin{equation}
\begin{aligned}
{\rm{cons.}}\{&P_{\rm{NER}}(x) \!\widetilde+\! P_{\rm{TR}}(x) \!\widetilde+\! P_{\rm{FER}}(x)\} \!=\! {P_{\rm{proto}}\left\{1, J \!+\! 1 \right\}},\\
{\rm{such \; that}} \quad &{\rm{gap}}_{i,j}>0, \;\, 0<i,j\le n,\\
&J\in \Big[{{\rm{max}}\left\lceil {P_{{\rm{sub}}m}} \right\rceil}, {{\rm{max}}\left\lceil {P_{\rm{SA}}} \right\rceil}\Big]\\
&{\rm{num}} \{ {\rm{gap}}_{i,j}=1 \} =1,\\
&{\rm{num}} \{ {\rm{gap}}_{i,j}=2 \}=1,\\
\end{aligned}
\label{ULA_fitting_strategy}
\end{equation}
where ${\rm{cons.}}\{P(x)\}$ stands for the consecutive polynomial of $P(x)$, ${\rm{max}}\left\lceil {P(x)} \right\rceil$ returns the maximum exponent value, ${\rm{num}}\{x\}$ returns the number of $x$, and $J$ is the positive interval that includes the consecutive range in DCA.

Clearly, the relationship between uDOF and $J$ is
\begin{equation}
{\rm {uDOF}}=2J+1.
\label{uDOF_and_J}
\end{equation}
Constraints in~(\ref{ULA_fitting_strategy}) are much easier to satisfy than constraint~(\ref{hole_free_FDCA}).

Note that (\ref{ULA_fitting_strategy}) is just an example. 

\subsection{Lower Bound of uDOF and Selection of $S_{\rm{t}}$}
\hfill \par
In ULA fitting, as we have analyzed, uDOF is dominated by $AP_{\rm{t}}$. Hence, the lower bound on the available uDOF is analyzed through $AP_{\rm{t}}$. We have mentioned that there are one SDCA and $n-1$ IDCAs within TR with period $S_{\rm{t}}$. Then, based on Proposition~\ref{proposition_coarray_number} and properties of the transfer layer, we can formulate a criteria for selecting $S_{\rm t}$.
\par
Generally speaking, $S_{\rm{t}}$ is determined according to $N_{\rm{base}}$, $N_{\rm{b}}$, $N_{\rm{addition}}$, and $N_{\rm{a}}$. Ignoring the coarrays in NER and FER, the maximum number of sensors within each period in TR is $N_{\rm{base}} \cdot N_{\rm{b}} + \sum _{i=1}^{N_{\rm{addition}}} N_{\rm{a}} (i) + 1$. Hence, if the interspace of the transfer layer is selected based on
\begin{equation}
S_{\rm{t}} \le N_{\rm{base}} \cdot N_{\rm{b}} + \sum _{i=1}^{N_{\rm{addition}}} N_{\rm{a}} (i)+1,
\label{S_t}
\end{equation}
a long ULA in TR can be guaranteed by properly selecting $S_{\rm{b}}$, $S_{\rm{a}}$, and gaps.
\par
Let us set $S_{\rm{t}} = N_{\rm{b}} + N_{\rm{a}}+z$, where $z \le 1$ is an integer. Further, considering total number of sensors $N$, we have
\begin{equation}
\left\{{\begin{array}{*{20}{l}}
N = N_{\rm{base}} \cdot N_{\rm{b}} + \sum _{i=1}^{N_{\rm{addition}}} N_{\rm{a}} (i)+N_{\rm{t}},\\
AP_{\rm{t }}= (N_{\rm{t}}-1)S_{\rm{t}}.
\end{array}} \right.
\label{AP_T}
\end{equation}
Substituting $S_{\rm{t}}$ into~(\ref{AP_T}), we obtain
\begin{equation}
AP_{\rm{t }}= -N_{\rm{t}}^2+(N+z+1)N_{\rm{t}}-N-z.
\label{AP_t_1}
\end{equation}
Maximizing~(\ref{AP_t_1}), yields the selection of $N_{\rm{t}}$, which is
\begin{equation}
N_{\rm{t }}= \frac{N+z+1}{2} .
\label{N_t}
\end{equation}
Considering the odevity of $N+z$,  $AP_{\rm{t}}^{\rm{max}}$ can be expressed as
\begin{equation}
AP_{\rm{t}}^{\rm{max}}=\frac{N^2+z^2+(2N-2)z-2N+\alpha}{4},
\label{AP_T_max}
\end{equation}
where $\alpha$ satisfies
\begin{equation}
\alpha =\left\{{\begin{array}{*{20}{l}}
0,&N+z {\rm{\ is\ odd}},\\
1,&N+z {\rm{\ is\ even}}.
\end{array}} \right.
\label{alpha}
\end{equation}
Omitting $\alpha$, which does not influence the monotonicity of $AP_{\rm{t}}^{\rm{max}}$,~(\ref{AP_T_max}) can be regarded as a function of $z$ which is monotonically increasing for $z\le 1$. Hence, $z=1$ can be selected, for example, to pursue a long uDOF. By selecting $z=1$, $AP_{\rm{t}}^{\rm{max}}$ can be written as
\begin{equation}
AP_{\rm{t}}^{\rm{max}}=\frac{N^2+\beta}{4},
\label{AP_T_max_final}
\end{equation}
where $\beta$ satisfies
\begin{equation}
\beta =\left\{{\begin{array}{*{20}{l}}
-1,&N {\rm{\ is\ odd}},\\
0,&N {\rm{\ is\ even}}.
\end{array}} \right.
\label{alpha}
\end{equation}

Therefore, according to~(\ref{AP_T_max_final}) for $z=1$, the selection criteria of $S_{\rm{t}}$ along with the corresponding lower bound on uDOF are
\begin{equation}
\left\{{\begin{array}{*{20}{l}}
S_{\rm{t}} = N_{\rm{base}} \cdot N_{\rm{b}} + \sum _{i=1}^{N_{\rm{addition}}} N_{\rm{a}} (i)+1,\\
{\rm{uDOF}}_{\rm{lower}}=\frac{N^2+\beta}{2}+1.
\end{array}} \right.
\label{lower_bound_uDOF}
\end{equation}
In fact, both $S_{\rm{t}} = N_{\rm{base}} \cdot N_{\rm{b}} + \sum _{i=1}^{N_{\rm{addition}}} N_{\rm{a}} (i)+1$ and $S_{\rm{t}} = N_{\rm{base}} \cdot N_{\rm{b}} + \sum _{i=1}^{N_{\rm{addition}}} N_{\rm{a}} (i)$ are good options.

For instance, the nested and MISC arrays~\cite{NESTED,MISC} both can be explained based on ULA fitting. The nested array is built based on one base layer consisted of only 1 ULA and a transfer layer also consisted of only 1 ULA, where $S_{\rm{t}} = N_{\rm{b}}+1$ is selected. In turns, the MISC array is built based on one base layer consisted of $2$ sub-ULAs, one transfer layer and one addition layer consisted of one sub-ULA with 2 sensors and interspace 1, $S_{\rm{t}} = N_{\rm{base}} \cdot N_{\rm{b}} + \sum _{i=1}^{N_{\rm{addition}}} N_{\rm{a}} (i)$ is selected.

\par
\section{SAs Design via ULA Fitting} \label{UF_example}
Before explaining the design procedure, as we have a lot of sub-ULAs to be arranged, more notations need to be declared first.

Sub-ULAs are differentiated by the layer they pertain to and the interspace. For example, the sub-ULAs that pertain to the base layer with interspace $3$ and addition layer with interspace $2$ can be expressed as $B_3$ and $A_2$, respectively, where subscripts indicate the interspaces. The transfer layer is expressed by $T$. The precedence of the sub-ULAs is shown by arrows. For example, the sequence corresponding to the SA shown in~Fig.~\ref{illustration_n2} can be expressed as
\begin{equation}
A_2 \rightarrow
B_3 \rightarrow
T \rightarrow
A_1 \rightarrow
B_3 \rightarrow
B_3.
\label{sequence_of_P3BASE3}
\end{equation}

\subsection{SA Design Procedure}
The procedure of designing SAs via ULA fitting can be summarized as follows.
\begin{enumerate}
\item Select the parameters of each layer and list target equation.
\item Determine the sequence of all sub-ULAs.
\item Find specific solution which satisfies the design criteria, and further analyze uDOF and parameter selection.
\item If no solution is obtained, then redo step 2 and select another sequence.
\end{enumerate}

\par
Now that all the necessary components have been declared, we give two examples of SA design to show how the ULA fitting works. Since the earlier mentioned MISC array already exhibits a solution when the base layer consisted of 2 sub-ULAs, we start here by designing an SA with the base layer consisted of 3 sub-ULAs to demonstrate the abilities of ULA fitting that are not matched in the existing literature of SAs. Note that in all the designs, the total number of sensors is denoted as $N$. Because of the space limitation, the solutions are presented directly.

\subsection{ULA Fitting for SAs Design Using Base Layer Consisted of 3 sub-ULAs (UF-3BL)}

\par
In the first example, we present the explanation of the design process to show how the ULA fitting works. Consider an SA that is built based on one base layer consisted of 3 sub-ULAs, one transfer layer, and two addition layers. The two addition layers consisted of 2-sensors each are selected with interspaces $1$ and $2$, respectively, to control weights $w(1)$ and $w(2)$. Therefore, in this case, the following relationships can be established
\begin{equation}
\left\{{\begin{array}{*{20}{l}}
N=3N_{\rm{b}}+N_{\rm{t}}+4,\\
S_{\rm{t}}=3N_{\rm{b}}+5.
\end{array}} \right.
\label{UF-3BLC1_parameter_relationship}
\end{equation}
The design problem can be expressed as
\begin{equation}
\begin{aligned}
{\rm{cons.}}\{P_{\rm{NER}}(x)\widetilde+&P_{\rm{TR}}(x)\widetilde+P_{\rm{FER}}(x)\}={P_{\rm{proto}}\left\{1 ,J+1 \right\}},\\
{\rm{such \, that}} \quad &{\rm{gap}}_{i,j}>2, \;\, 0<i,j\le n,\\
&J\in \Big[{{\rm{max}}\left\lceil {P_{{\rm{sub}}m}} \right\rceil}, {{\rm{max}}\left\lceil {P_{\rm{SA}}} \right\rceil}\Big].\\
\end{aligned}
\label{Strategy_of_UF-3BLC1}
\end{equation}
Moreover, we get $6$ sub-ULAs to be arranged, where the sequence is selected as
\begin{equation}
B_3 \rightarrow
A_1 \rightarrow
T \rightarrow
B_3 \rightarrow
A_2 \rightarrow
B_3.
\label{sequence_of_UF-3BLC1}
\end{equation}
Note that, based on the dual property, sequence~(\ref{sequence_of_UF-3BLC1}) leads to the same solution as the following permuted sequence
\begin{equation}
B_3 \rightarrow
A_2 \rightarrow
B_3 \rightarrow
T \rightarrow
A_1 \rightarrow
B_3.
\end{equation}
Hence, the significance of the dual property is to reduce the possible permutations greatly. Considering~(\ref{polynomial_NER_TR_FER}) and~(\ref{Strategy_of_UF-3BLC1}), a particular solution for this case is obtained as
\begin{equation}
\left\{{\begin{array}{*{20}{l}}
{\rm{gap}}_{1,2}=4,\\
{\rm{gap}}_{2,3}=2+3N_{\rm{b}},\\
{\rm{gap}}_{3,4}=3,\\
{\rm{gap}}_{4,5}=4,\\
{\rm{gap}}_{5,6}=3.\\
\end{array}} \right.
\label{solution_UF-3BLC1}
\end{equation}
With~(\ref{solution_UF-3BLC1}), $P_{\rm{NER}}(x)$, $P_{\rm{TR}}(x)$ and $P_{\rm{FER}}(x)$ are expressed as
\begin{equation} 
\begin{aligned}
P_{\rm{NER}}(x) &\quad = \{x^0+x^1+x^2\}\\
&\quad  \widetilde+x^9  \times P_{\rm{proto}}\{3,2N_{\rm{b}}-1\}\\
&\quad  \widetilde+\{x^0+x^3+x^4+x^5+x^6\} \times P_{\rm{proto}}\{3,N_{\rm{b}}\},\\
\end{aligned}
\label{NER_3_UF-3BLC1}
\end{equation}
\begin{equation} 
\begin{aligned}
P_{\rm{TR}}(x) &\quad= x^0\widetilde+x^3\widetilde\times P_{\rm{proto}} \{3,N_{\rm{b}}\}\\
&\quad  \widetilde+\underbrace{x^{3N_{\rm{b}}+2}+\cdots+x^{3N_{\rm{b}}N_{\rm{t}}+5N_{\rm{t}}-1}}_{\text{consecutive}}\\
&\quad  \widetilde+x^{3N_{\rm{b}}+7} \times P_{\rm{proto}}\{3,N_{\rm{b}}\} \times P_{\rm{proto}}\{3N_{\rm{b}}+5,N_{\rm{t}}\}\\
&\quad  \widetilde+x^{3N_{\rm{b}}+9} \times P_{\rm{proto}}\{3,N_{\rm{b}}\} \times P_{\rm{proto}}\{3N_{\rm{b}}+5,N_{\rm{t}}\},\\
\end{aligned}
\label{TR3_UF-3BLC1}
\end{equation}
and
\begin{equation} 
\begin{aligned}
P_{\rm{FER}}(x) &\quad = x^{3N_{\rm{b}}N_{\rm{t}} + 5N_{\rm{t}}+5} \times P_{\rm{proto}} \{3,2N_{\rm{b}}-1\} \\
&  \widetilde +\{x^0+x^2\}\times x^{3N_{\rm{b}}N_{\rm{t}}+5N_{\rm{t}}+3N_{\rm{b}}+6}\times P_{\rm{proto}}\{3,N_{\rm{b}}\}\\
&  \widetilde +x^{3N_{\rm{b}}N_{\rm{t}}+5N_{\rm{t}}+3N_{\rm{b}}+11}\times   P_{\rm{proto}}\{3,2N_{\rm{b}}-1\}\\
&  \widetilde +\{x^0+x^1\} \times x^{3N_{\rm{b}}N_{\rm{t}}+5N_{\rm{t}}}\times P_{\rm{proto}}\{3,N_{\rm{b}}\}\\
&  \widetilde +\{x^0+x^1+x^2+x^3\}\times x^{3N_{\rm{b}}N_{\rm{t}}+5N_{\rm{t}}+3N_{\rm{b}}+1}\\
&  \widetilde +\{x^0+x^1\}\times x^{3N_{\rm{b}}N_{\rm{t}}+5N_{\rm{t}}+3N_{\rm{b}}+6}\times P_{\rm{proto}}\{3,N_{\rm{b}}\},\\
\end{aligned}
\label{FER_3_UF-3BLC1}
\end{equation}
which leads to the following expression
\begin{equation}
\begin{aligned}
{\rm{cons.}}\{&P_{\rm{NER}}(x)\widetilde+P_{\rm{TR}}(x)\widetilde+P_{\rm{FER}}(x)\}\\
&={P_{\rm{proto}}\left\{1,3N_{\rm{b}}N_{\rm{t}}+5N_{\rm{t}}+3N_{\rm{b}} \right\}}.\\
\end{aligned}
\label{J_UF-3BLC1}
\end{equation}
Comparing~(\ref{Strategy_of_UF-3BLC1}) and~(\ref{J_UF-3BLC1}), we have
\begin{equation}
J_{\rm{UF-3BL}}=3N_{\rm{b}}N_{\rm{t}}+5N_{\rm{t}}+3N_{\rm{b}}-1.
\label{J_UF-3BLC1_1}
\end{equation}
Considering the fact that $N_{\rm{b}}$ and $N_{\rm{t}}$ are integers and the total number of sensors $N$ is fixed, the following optimal selection of $N_{\rm{b}}$ and $N_{\rm{t}}$ is obtained by maximizing~(\ref{J_UF-3BLC1_1})
\begin{equation}
\left\{{\begin{array}{*{20}{l}}
N_{\rm{b}} = \lfloor \frac{N-5}{6}\rfloor,N\ge17,\\
N_{\rm{t}} = N-3N_{\rm{b}}-4,
\end{array}} \right.
\label{UF-3BLC1_1_parameter_selection}
\end{equation}
where $\lfloor \cdot \rfloor$ is the floor operation.
The final uDOF can then be expressed as
\begin{equation}
{\rm{uDOF}}=
\left\{{\begin{array}{*{20}{l}}
\frac {N^2}{2} +2N - 11,      &N\%6=0,4\\
\frac {N^2}{2} +2N - 9.5,    &N\%6=1,3\\
\frac {N^2}{2} +2N - 9,      &N\%6=2,\\
\frac {N^2}{2} +2N - 13.5,    &N\%6=5,\\
\end{array}} \right.
,N\ge 17,
\label{UF-3BLC1_1_uDOF}
\end{equation}
where $\%$ stands for the remainder.
The closed-form expressions for the proposed UF-3BL can be summarized as
\begin{equation} 
\left\{{\begin{array}{*{20}{l}}
&{\text{sub-ULA1}}:  \left\{0,3,N_{\rm{b}}\right\},\\
&{\text{sub-ULA2}}:  \left\{3N_{\rm{b}}+1,1,2\right\}, \\
&{\text{sub-ULA3}}:  \left\{6N_{\rm{b}}+4,3N_{\rm{b}}+5,N_{\rm{t}}\right\}, \\
&{\text{sub-ULA4}}:  \left\{3N_{\rm{t}}N_{\rm{b}}+5N_{\rm{t}}+3N_{\rm{b}}+2,3,N_{\rm{b}}\right\}, \\
&{\text{sub-ULA5}}:  \left\{3N_{\rm{t}}N_{\rm{b}}+5N_{\rm{t}}+6N_{\rm{b}}+3,2,2 \right\}, \\
&{\text{sub-ULA6}}:  \left\{3N_{\rm{t}}N_{\rm{b}}+5N_{\rm{t}}+6N_{\rm{b}}+8,3,N_{\rm{b}}\right\}. \\
\end{array}} \right.
\label{UF-3BLC1_structure}
\end{equation}
Note that, in~(\ref{UF-3BLC1_1_parameter_selection}) and~(\ref{UF-3BLC1_1_uDOF}), $N\ge17$ is required to obtain the optimal uDOF.
\par
Further, the weight function of the proposed UF-3BL for coarray indexes $1$, $2$, and $3$, i.e., the weights $w(1)$, $w(2)$, and $w(3)$, can be easily found to be
\begin{equation}
w(1)=1, \quad w(2)=1, \quad w(3)=3N_{\rm{b}}-1.
\label{WF_3_base_c2_advanced}
\end{equation}
Here $w(1)$ stems from $A_1(1)$ and $w(2)$ is generated from $A_2(2)$, while $w(3)$ is contributed by the three sub-ULAs of the base layer ($3\times(N_{\rm{b}}$-1)) and 2 gaps (${\rm{gap}}_{3,4}$ and ${\rm{gap}}_{5,6}$).

\subsection{ULA Fitting for SAs Design Using Base Layer Consisted of 4 sub-ULAs (UF-4BL)}
\par
We further consider the case of using the base layer consisted of 4 sub-ULAs to design SA. Additionally, we use 3 addition layers each consisted of 2-sensors with interspaces $1$, $2$, and $3$ to control $w(1)$, $w(2)$, and $w(3)$. In this case, $8$ sub-ULAs are considered with the following sequence
\begin{equation}
\begin{aligned}
&A_3 \rightarrow
B_4 \rightarrow
A_1 \rightarrow
B_4 \rightarrow T \rightarrow
B_4 \rightarrow
A_2 \rightarrow
B_4,
\end{aligned}
\end{equation}
and, the following relationships are satisfied
\begin{equation}
\left\{{\begin{array}{*{20}{l}}
N=4N_{\rm{b}}+N_{\rm{t}}+6,\\
S_{\rm{t}}=4N_{\rm{b}}+7.
\end{array}} \right.
\label{UF-4BLC1_parameter_relationship}
\end{equation}

The corresponding design problem inherits~(\ref{Strategy_of_UF-3BLC1}), and one possible solution can be obtained as
\begin{equation}
\left\{{\begin{array}{*{20}{l}}
{\rm{gap}}_{1,2}=4,\\{\rm{gap}}_{2,3}=5,\\
{\rm{gap}}_{3,4}=6,\\{\rm{gap}}_{4,5}=8,\\
{\rm{gap}}_{5,6}=7,\\{\rm{gap}}_{6,7}=3,\\
{\rm{gap}}_{7,8}=5,\\
\end{array}} \right.
\label{solution_UF-4BLC1}
\end{equation}
with corresponding $P_{\rm{NER}}(x)$, $P_{\rm{TR}}(x)$, and $P_{\rm{FER}}(x)$ given as
\begin{equation} 
\begin{aligned}
P_{\rm{NER}}(x) &\quad= \{x^0+x^1+x^2+x^3\}\\
&\widetilde+\{x^0+x^1+x^3+x^4\}\times x^{4N_{\rm{b}}+5}\\
&\widetilde+(x^{10}+x^{12})\times P_{\rm{proto}}\{4,2N_{\rm{b}}-1\}\\
&\widetilde+\{x^0+x^3+x^4+x^5+x^6+x^7+x^{4N_{\rm{b}}+12} \\
&+ x^{4N_{\rm{b}}+15} \} \times P_{\rm{proto}} \{4, N_{\rm{b}} \},\\
\end{aligned}
\label{NER_UF-4BLC1}
\end{equation}
\begin{equation} 
\begin{aligned}
&P_{\rm{TR}}(x)=x^0\widetilde+(x^6+x^7+x^8+x^{10}+x^{11})\times x^{4N_{\rm{b}}}\\
&\quad \quad \widetilde+(x^7+x^8)\times P_{\rm{proto}}\{4,N_{\rm{b}}\}\\
&\quad \quad \widetilde+\underbrace{x^{4N_{\rm{b}}+13}+\cdots+x^{4N_{\rm{t}}N_{\rm{b}}+7N_{\rm{t}}-1}}_{\text{consecutive}}\\
&\quad \quad \widetilde+(x^6+x^9)\times x^{4N_{\rm{t}}N_{\rm{b}}+7N_{\rm{t}}}P_{\rm{proto}}\{4,N_{\rm{b}}\}\\
&\quad \quad \widetilde+(x^1+x^2+x^3+x^4+x^5)\times x^{4N_{\rm{t}}N_{\rm{b}}+7N_{\rm{t}}+4N_{\rm{b}}},\\
\end{aligned}
\label{TR_UF-4BLC1}
\end{equation}
and
\begin{equation} 
\begin{aligned}
&P_{\rm{FER}}(x)= (x^{9}+x^{10}+x^{11}+x^{12})\times x^{4N_{\rm{b}}N_{\rm{t}}+7N_{\rm{t}}+4N_{\rm{b}}} \\
&\quad x^{4N_{\rm{b}}N_{\rm{t}}+7N_{\rm{t}} - 4N_{\rm{b}} + 8}\times P_{\rm{proto}}\{4,2N_{\rm{b}}-1\}\\
&\quad \widetilde +(x^{14}+x^{16})\times x^{4N_{\rm{b}}N_{\rm{t}}+7N_{\rm{t}}}\times P_{\rm{proto}}\{4,2N_{\rm{b}}-1\} \\
&\quad \widetilde +x^{4N_{\rm{b}}N_{\rm{t}}+7N_{\rm{t}} + 4N_{\rm{b}}+22}\times P_{\rm{proto}}\{4,2N_{\rm{b}}-1\} \\
&\quad \widetilde +(x^7+x^9+x^{10}+x^{11})\times x^{4N_{\rm{b}}N_{\rm{t}}+7N_{\rm{t}}}\times P_{\rm{proto}}\{4,N_{\rm{b}}\} \\
&\quad \widetilde +\big[(x^{15}+x^{16}+x^{17}+x^{19})\times x^{4N_{\rm{b}}N_{\rm{t}}+7N_{\rm{t}}+4N_{\rm{b}}}\\
&\quad\quad \widetilde +(x^{15}+x^{18}+x^{22}+x^{25})\times x^{4N_{\rm{b}}N_{\rm{t}}+7N_{\rm{t}}+8N_{\rm{b}}}\big]\\
&\quad \widetilde \times P_{\rm{proto}}\{4,N_{\rm{b}}\},\\
\end{aligned}
\label{FER_UF-4BLC1}
\end{equation}
leading to
\begin{equation}
\begin{aligned}
{\rm{cons.}}\{&P_{\rm{NER}}(x)\widetilde+P_{\rm{TR}}(x)\widetilde+P_{\rm{FER}}(x)\}\\
&={P_{\rm{proto}}\left\{1,4N_{\rm{b}}N_{\rm{t}}+7N_{\rm{t}}+4N_{\rm{b}}+13 \right\}},\\
\end{aligned}
\label{J_UF-4BLC1}
\end{equation}
and, thus,
\begin{equation}
J_{\rm{UF-4BL}}=4N_{\rm{b}}N_{\rm{t}}+7N_{\rm{t}}+4N_{\rm{b}}+12.
\label{J_UF-4BLC1_1}
\end{equation}
Similar to~(\ref{UF-3BLC1_1_parameter_selection}), the optimal solution for parameters can be further obtained as
\begin{equation}
\left\{{\begin{array}{*{20}{l}}
N_{\rm{b}} = \lfloor \frac{N-8}{8}\rfloor,N\ge32,\\
N_{\rm{t}} = N-4N_{\rm{b}}-6.
\end{array}} \right.
\label{UF-4BLC1_1_parameter_selection}
\end{equation}
Following~(\ref{UF-4BLC1_parameter_relationship}),~(\ref{J_UF-4BLC1_1}) and~(\ref{UF-4BLC1_1_parameter_selection}), the final $\rm{uDOF}$ is summarized as
\begin{equation}
{\rm{uDOF}}=
\left\{{\begin{array}{*{20}{l}}
\frac {N^2}{2} +2N + 5,      &N\%6=0,\\
\frac {N^2}{2} +2N + 13.5,    &N\%6=1,7,\\
\frac {N^2}{2} +2N + 11,      &N\%6=2,6,\\
\frac {N^2}{2} +2N + 12.5,    &N\%6=3,5,\\
\frac {N^2}{2} +2N + 13,    &N\%6=4.\\
\end{array}} \right.
N\ge 32.
\label{UF-4BLC1_1_uDOF}
\end{equation}
The structure of the proposed UF-4BL can be summarized in closed-form as
\begin{equation} 
\left\{{\begin{array}{*{20}{l}}
&{\text{sub-ULA1}}:  \left\{0,3,2\right\},\\
&{\text{sub-ULA2}}:  \left\{7,4,N_{\rm{b}}\right\}, \\
&{\text{sub-ULA3}}:  \left\{4N_{\rm{b}}+8,1,2\right\}, \\
&{\text{sub-ULA4}}:  \left\{4N_{\rm{b}}+15,4,N_{\rm{b}}\right\}, \\
&{\text{sub-ULA5}}:  \left\{8N_{\rm{b}}+19,4N_{\rm{b}}+7,N_{\rm{t}} \right\} \\
&{\text{sub-ULA6}}:  \left\{4N_{\rm{t}}N_{\rm{b}}+7N_{\rm{t}}+4N_{\rm{b}}+19,4,N_{\rm{b}}\right\}, \\
&{\text{sub-ULA7}}:  \left\{4N_{\rm{t}}N_{\rm{b}}+7N_{\rm{t}}+8N_{\rm{b}}+18,2,2\right\}, \\
&{\text{sub-ULA8}}:  \left\{4N_{\rm{t}}N_{\rm{b}}+7N_{\rm{t}}+8N_{\rm{b}}+25,4,N_{\rm{b}}\right\}, \\
\end{array}} \right.
\label{structure_UF-4BLC1}
\end{equation}
with
\begin{equation}
	w(1)=1, \;\, w(2)=1, \;\, w(3)=2, \;\, w(4)=4N_{\rm{b}}-3.
\end{equation}

\par
Note that, for pursuing the optimal uDOF, both UF-3BL and UF-4BL pose a requirement to the total number of sensors $N$. When $N$ is less than required, UF-3BL and UF-4BL are still valid designs, but no longer satisfy~(\ref{UF-3BLC1_1_uDOF}) and~(\ref{UF-4BLC1_1_uDOF}).

\section{Numerical Experiments}\label{NE}
In this section, numerical experiments are used to verify the superiority of the proposed SA geometries in high mutual coupling environment. SA structures considered here are nested array, coprime array, SNA, ANA, MISC, and two structures proposed in Section~\ref{UF_example}. The performance is evaluated in terms of coupling leakage, spatial efficiency, uDOF, DOAs identifiability, and root-mean-square error (RMSE). The spatial efficiency is defined as
\begin{equation}
	{\rm{Spatial \ Efficiency}}=\frac {J_{\rm{SA}}}{AP_{\rm{SA}}}.
	\label{spatial_efficiency}
\end{equation}
The RMSE is given as
\begin{equation}
{\rm{RMSE}}=\sqrt{\frac{1}{PQ} \sum^{P}_{p=1}\sum^{Q}_{q=1}({\hat{\theta}_q^p}-{\theta {_q}})},
\end{equation}
where $P$ is the number of trials and ${\hat{\theta}_q^p}$ represents the estimation of ${\theta{_q}}$ in $p{\rm{th}}$ trial.
For all the SA geometries tested, DOAs are computed based on  spatial smoothing MUSIC algorithm~\cite{NESTED,spatial_smoothing}.

\subsection{uDOF, Spatial Efficiency and Coupling Leakage}
\begin{figure}[htp]
\quad\subfloat[uDOF versus number of sensors.]{\includegraphics[width=0.9\columnwidth,height=5.3cm]{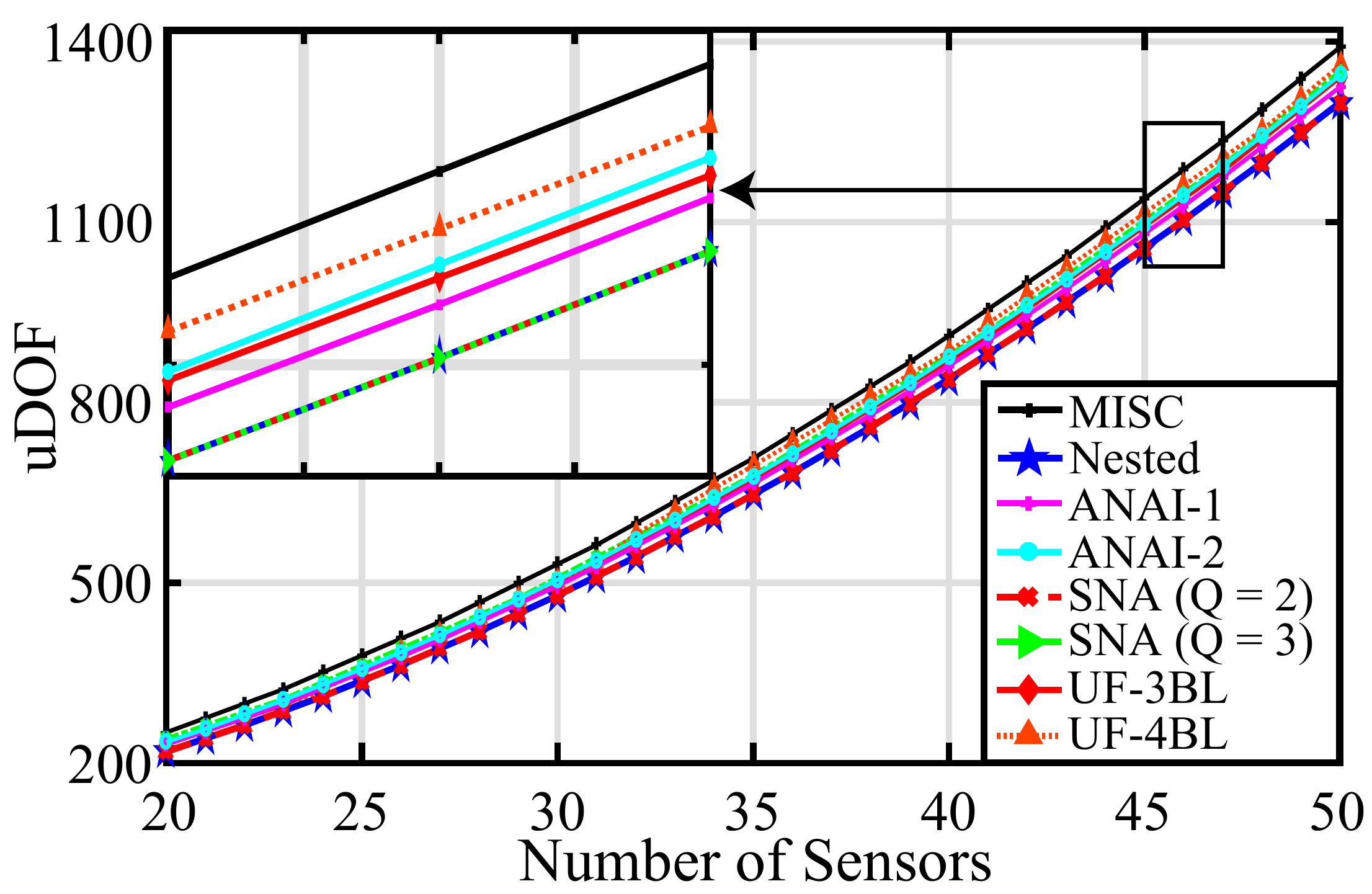}}\\
\centering
\subfloat[Spatial efficiency versus number of sensors.]{\includegraphics[width=0.9\columnwidth,height=5.0cm]{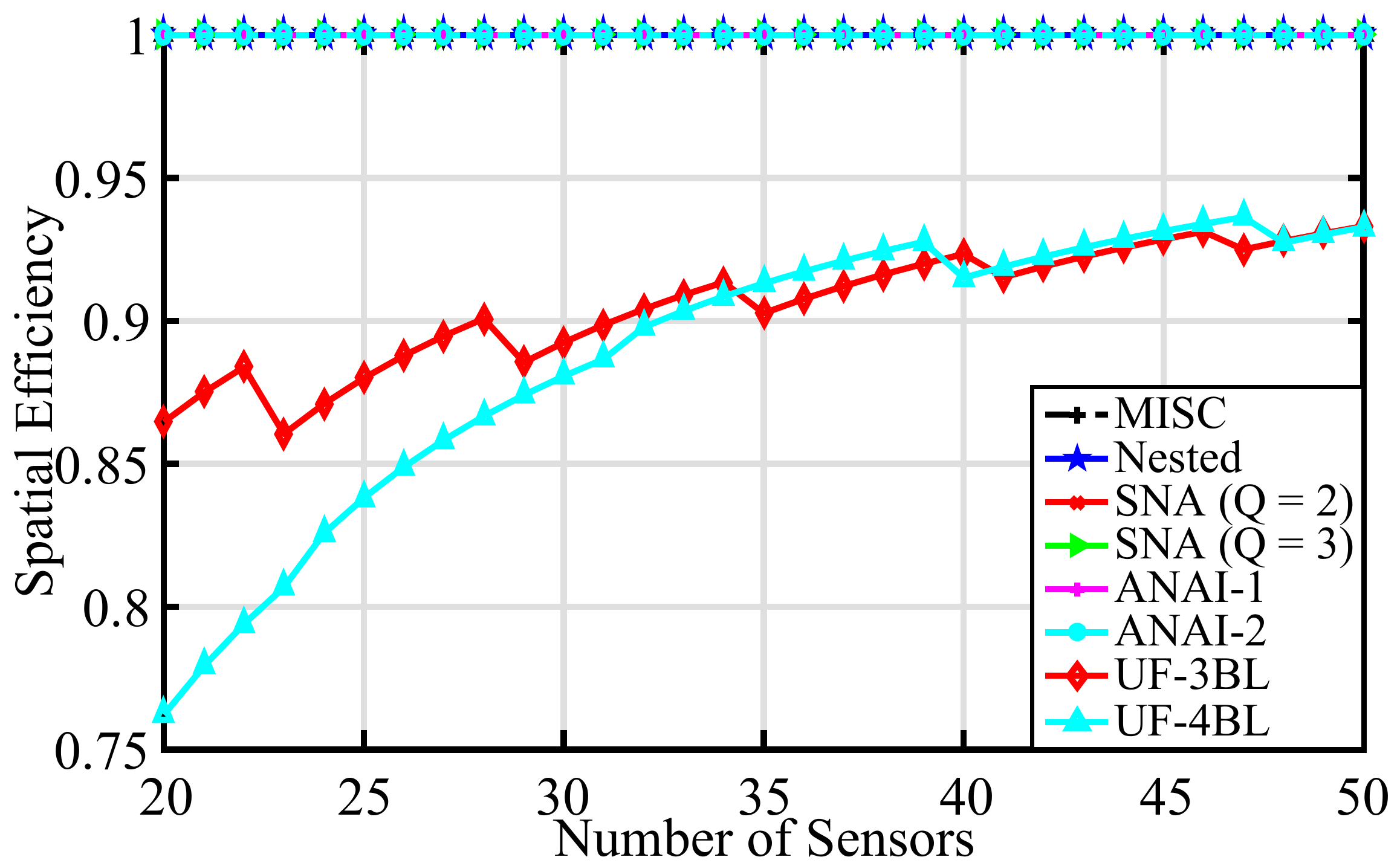}}\\
\centering
\subfloat[Coupling leakage versus number of sensors.]{\includegraphics[width=0.95\columnwidth,height=5.3cm]{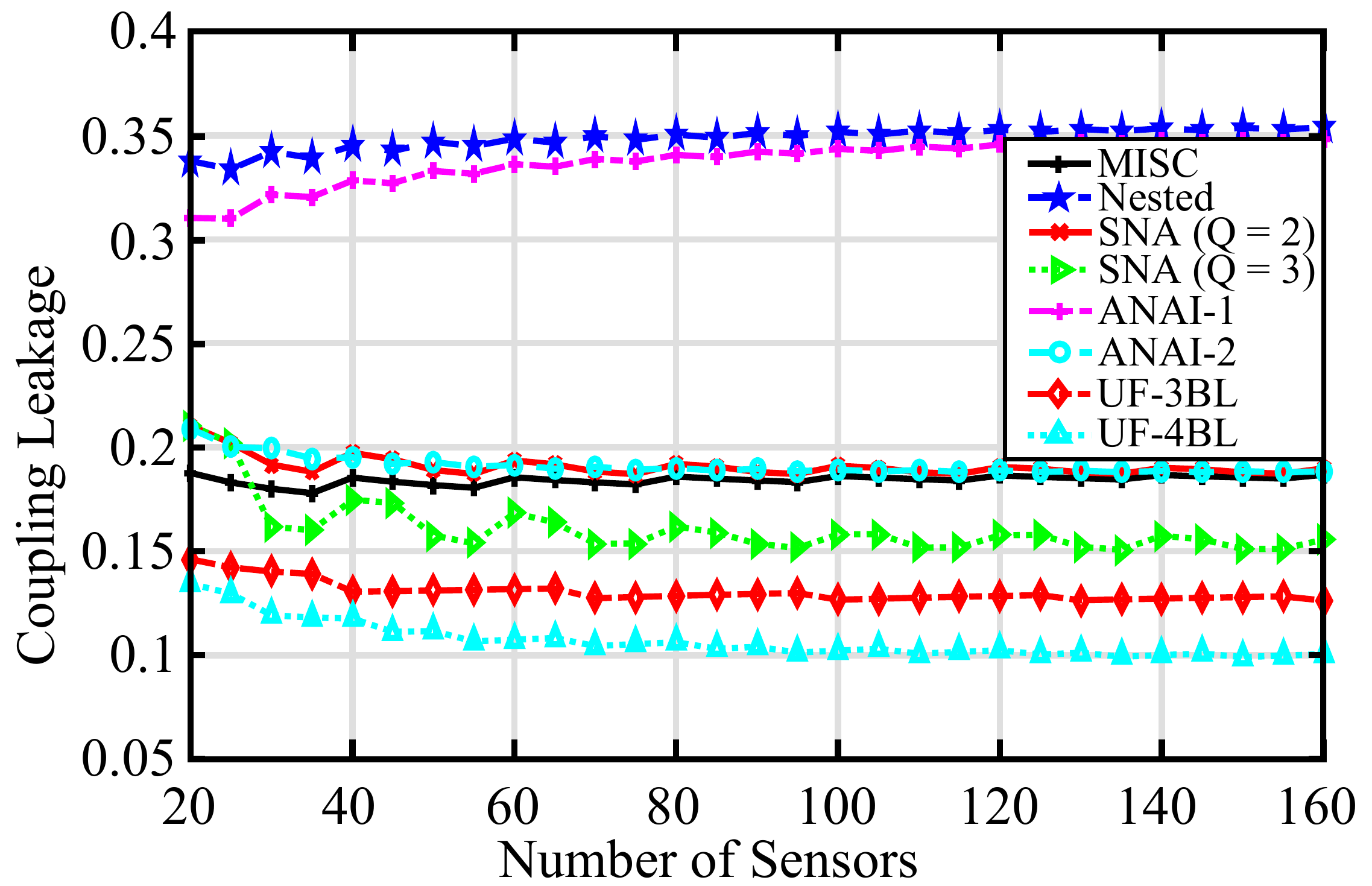}}
\caption{Illustration of uDOF, spatial efficiency, and coupling leakage.}
\label{ex_1}
\end{figure}

In the first example, we investigate the uDOF, spatial efficiency, and coupling leakage of the SA structures tested. Fig.~\ref{ex_1}~(a) shows the uDOFs of all the SA structures tested versus the number of array elements. Among the SA structures, the proposed UF-3BL and UF-4BL possess lower uDOF than MISC, but higher than the nested array and SNA. Fig.~\ref{ex_1}~(b) depicts the spatial efficiency of all the SA structures tested. For number of sensors larger than $35$, the spatial efficiency is larger than $90\%$. Moreover, as the number of sensors increases, the spatial efficiency also increases.

The coupling leakage versus number of sensors is presented in~Fig.~\ref{ex_1}~(c). It can be seen that the proposed UF-3BL and UF-4BL provide significantly reduced mutual coupling in comparison with SNA, MISC, and ANAI-2. The weight functions for small coarray indexes of relevant structures are also summarized in Table~\ref{weight_function_summarize}.
\begin{table}[htp]
\centering
\caption{Weight functions of SA structures tested}
\renewcommand{\arraystretch}{0.8}
\label{weight_function_summarize}

    \begin{tabular}{lcccc}
    \toprule
    Geometries                                      & w(1)      & w(2)     &w(3)       \\
    \midrule
    Nested                                          &$N_1$-1    & $N_1$-2  &$N_1$-3     \\
    \midrule
    Coprime                                         &2          & 2       &2        \\
    \midrule
    SNA ($Q=2$)                            &1 or 2          & ${N_1}-1$ or ${N_1} -3$      &1 or 3 or 4        \\
    \midrule
    SNA ($Q\ge3$)                          &2          & $\frac{N_1}{2}$ or $\frac{N_1}{2} -1$ or $\frac{N_1}{2} +1$      &5        \\
    \midrule
    ANAI-1                                          &$\frac{M}{2}$          &$\frac{M}{2}$       &$\frac{M}{2}$        \\
    \midrule
    ANAI-2                                          &2          &$\frac{M}{2}$       &2        \\
    \midrule
    MISC                                            &1         & $2\lfloor \frac{N}{4}\rfloor -2$      &1 or 2       \\
    \midrule
    UF-3BL                                        &1         & 1      &$3N_{\rm{b}}-1$       \\
    \midrule
    UF-4BL                                        &1         & 1      &2       \\
    \bottomrule
    \end{tabular}
\end{table}

\subsection{Target Identifiability and RMSE Performance}
\begin{figure}[htp]
\centering
\centerline{\includegraphics[width=1\columnwidth,height=7.0cm]{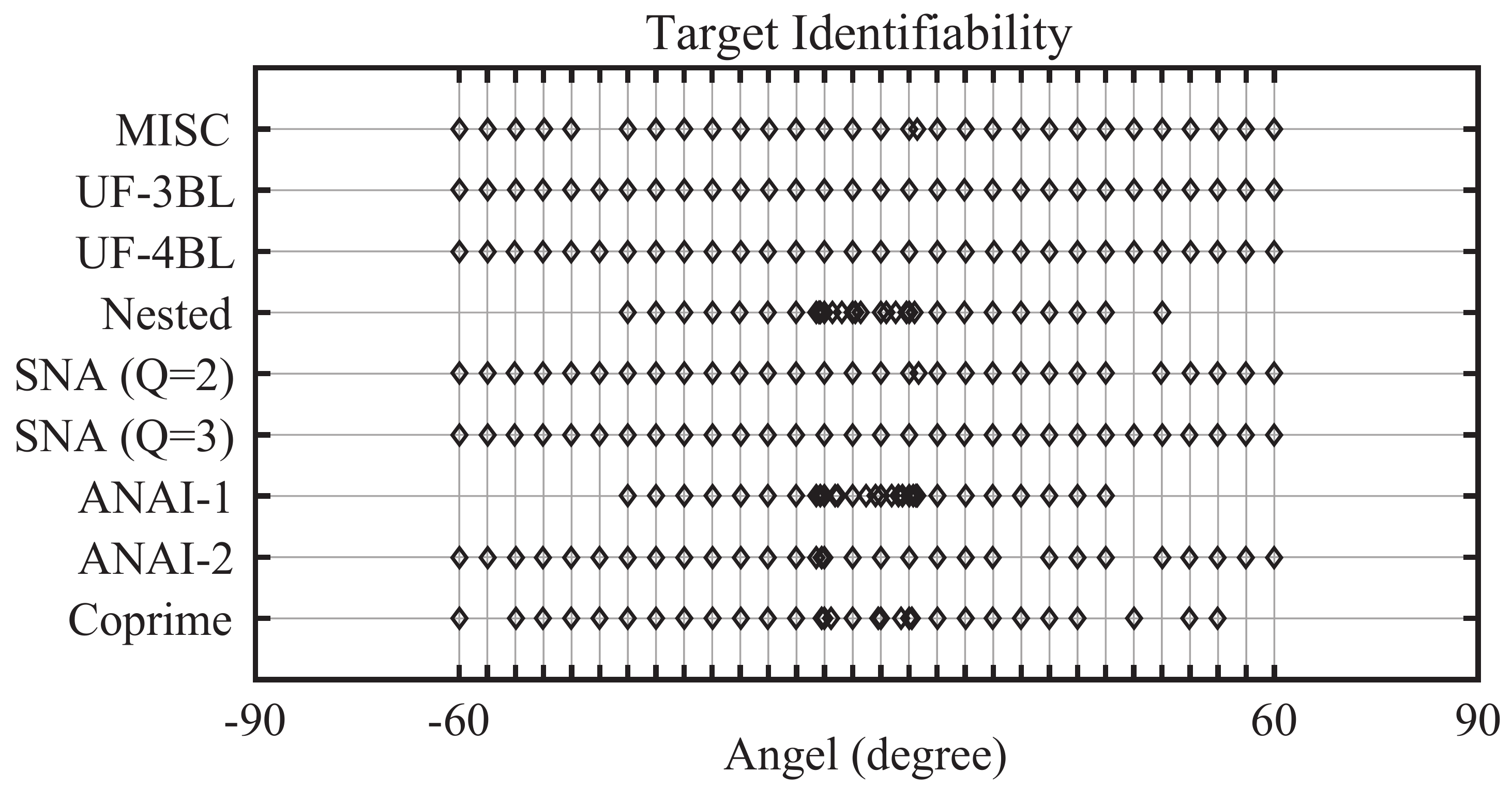}}
\caption{Target identification with 35 sensors, $|c_1|$=0.5}
\label{ex2_Tar}
\end{figure}

In our second example, we analyze the target identifiability and the RMSE performance for different conditions. First, the target identifiability in severe mutual coupling environment is shown in~Fig.~\ref{ex2_Tar}. Herein, $c_1=0.5e^{j\pi/3}$ and $c_i=c_1e^{-j(i-1)\pi/8}/i, \, i=2,\dots,100$, and SNR = 0~dB. All the SA structures possess $35$ sensors and $30$ targets which are uniformly distributed in azimuth in the interval from $-60^\circ$ to $60^\circ$. It can be seen from~Fig.~\ref{ex2_Tar} that the proposed UF-3BL, UF-4BL, and SNA ($Q=3$) can identify all the targets successfully, while other SA structures tested have missed and spurious targets.

\par

\begin{figure}[htp]
\centering
\centerline{\includegraphics[width=1.05\columnwidth,height=7cm]{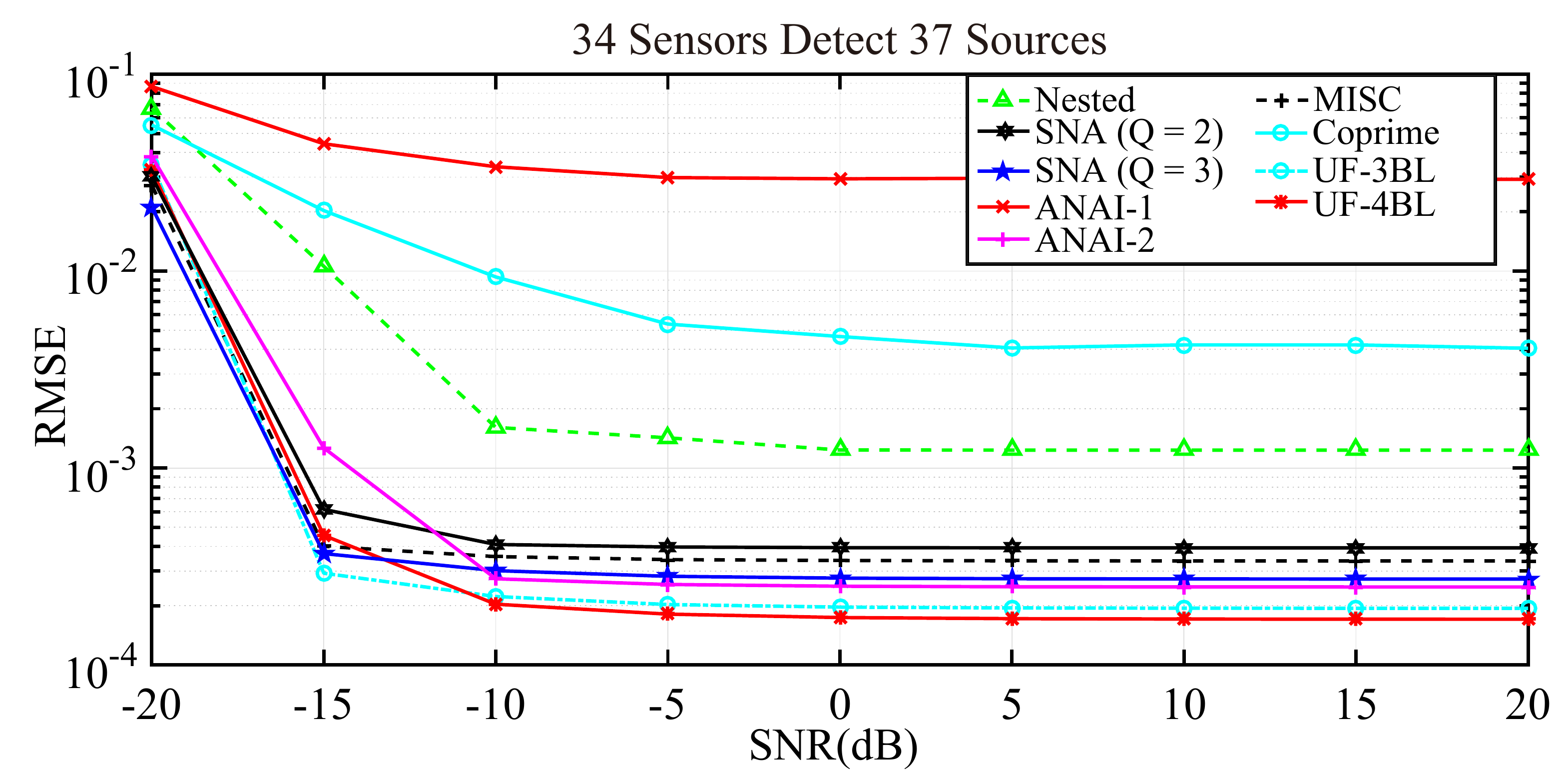}}
\caption{Illustration of RMSE versus SNR with 34 sensors.}
\label{ex2_RMSE_c1=03}
\end{figure}

Second, the RMSE performance versus SNR is presented. In this case, we consider $37$ targets within interval $\left[-60^\circ,60^\circ \right]$, $c_1=0.3e^{j\pi/3}$, $c_i=c_1e^{-j(i-1)\pi/8}/i, \, i=2,\dots,100$, and all SAs are composed of $34$ sensors. One can see from~Fig.~\ref{ex2_RMSE_c1=03} that the proposed UF-4BL performs the best when SNR is larger than -10 dB, while the proposed UF-3BL performs slightly worse than the proposed UF-4BL in high SNR environment. However, both the proposed UF-4BL and UF-3BL have a considerable improvement compared to the other SA structures tested.

\par
\begin{figure}[htp]
\centering
\centerline{\includegraphics[width=1.05\columnwidth,height=7cm]{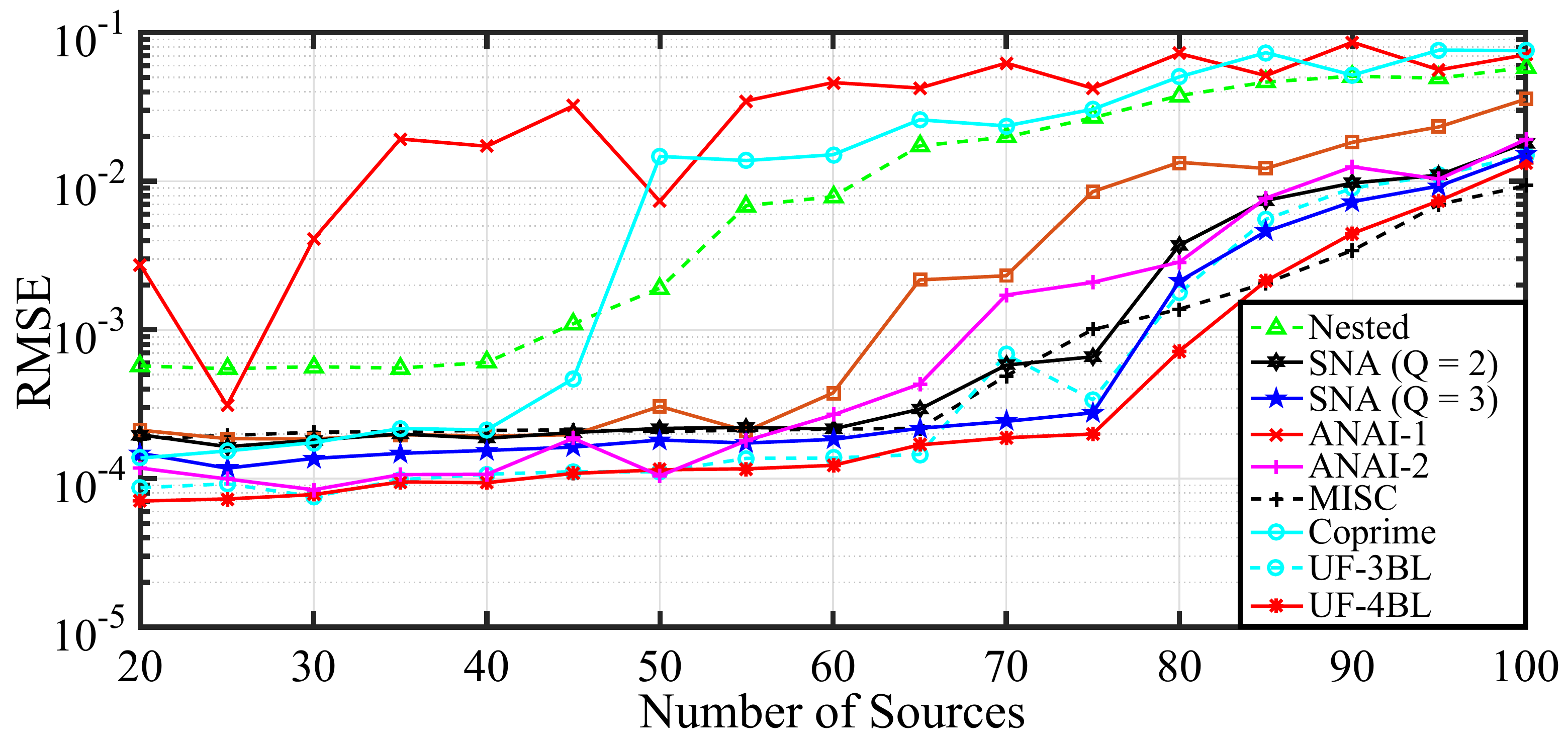}}
\caption{RMSE versus number of sources. All SA structures are composed of 44 sensors.}
\label{ex2_source_switch}
\end{figure}

Third, the RMSE performance versus the number of sources is studied. In this case, $c_1=0.3e^{j\pi/3}$ and $c_i=c_1e^{-j(i-1)\pi/8}/i, \, i=2,\dots,100$ are considered. The number of sources {\color{red}varies} from 20 to 100 and all the SA structures tested are composed of $44$ sensors. As illustrated in Fig.~\ref{ex2_source_switch}, the proposed UF-4BL and UF-3BL both perform well when the number of sources is smaller than $70$. However, the performance of the proposed UF-3BL deteriorates when the number of sources is larger than 70. When the number of sources is larger than 90, MISC provides the best performance. This is because MISC provides a larger uDOF than that of the proposed UF-3BL and UF-4BL.

\begin{figure}[htp]
	\centering
	\centerline{\includegraphics[width=1.05\columnwidth,height=6.7cm]{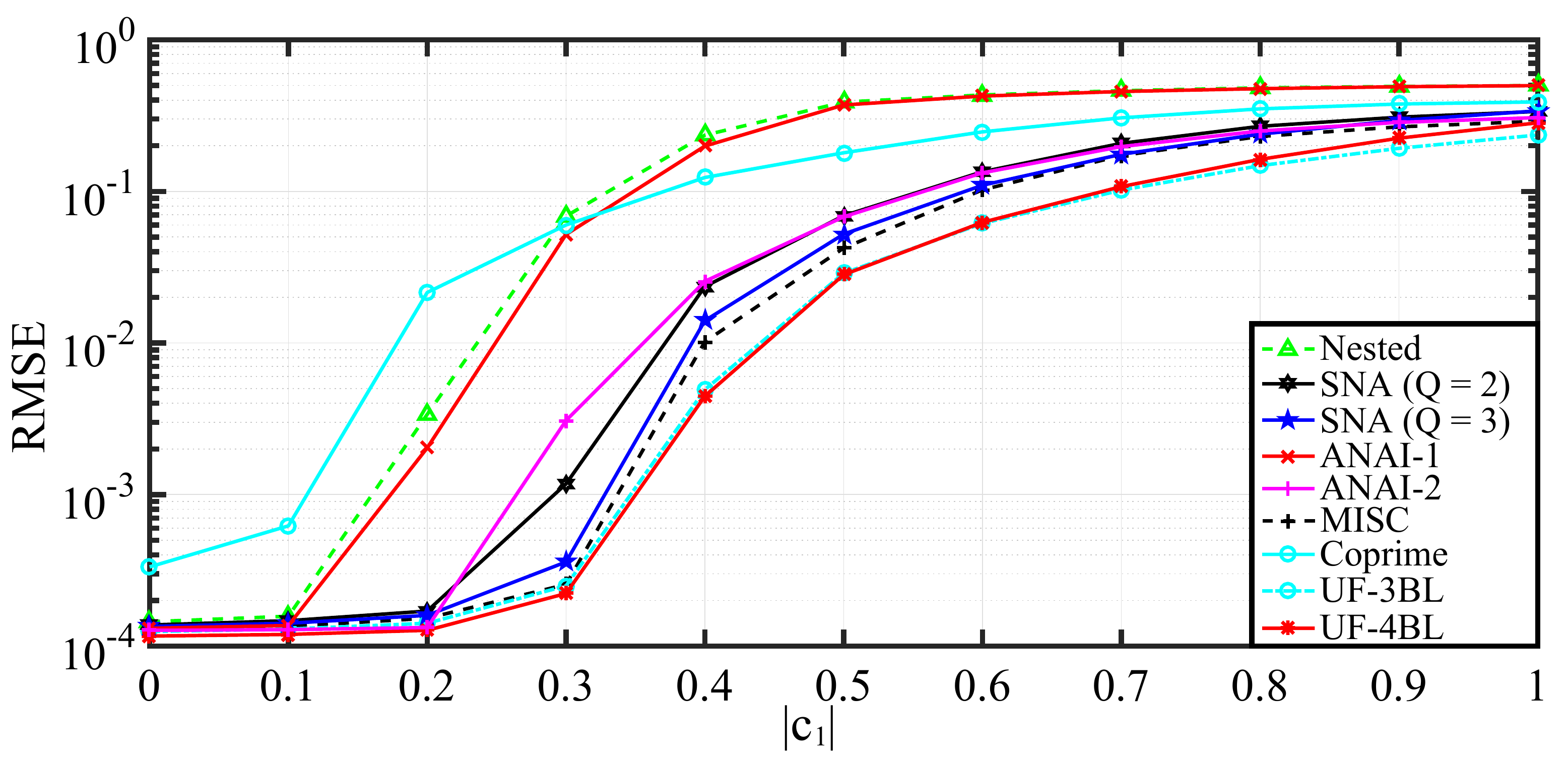}}
	\caption{Illustration of RMSE versus $|c_1|$, SNR = 0~dB. All SA structures are composed of 44 sensors.}
	\label{ex2_c1_switch}
\end{figure}

\par
Finally, the RMSE performance versus $|c_1|$ is presented. In this example, $50$ sources are uniformly distributed in the interval $\left[-60^\circ,60^\circ \right]$, and all the SA structures tested are composed of $44$ sensors. It is apparent from~Fig.~\ref{ex2_c1_switch} that when $|c_1|\le0.2$, i.e., in low mutual coupling environment, the RMSE is mainly determined by the uDOF. However, when $|c_1|\ge0.2$, the two proposed SA structures show better performance than the other SA structures tested.

\section{Conclusion}\label{conclusion}
In this paper, an SA design principle, ULA fitting, is established. The ULA fitting enables using pseudo polynomial equations corresponding to arrays to design SAs with closed-form expressions, large uDOF, and low mutual coupling. 
Two examples of designing SA structures based on the proposed ULA fitting are given. Numerical examples show the effectiveness of these SA structures. Although the general principle of ULA fitting has been introduced, we limited our detailed study by considering currently only one base layer. Thus, the proposed specific examples of SA structures (UF-3BL and UF-4BL) are limited in terms of uDOF. 
Future work includes the use of ULA fitting to design SA structures using base layer with even larger interspace and investigating ULA fitting with more base layers to design SA structures with further improved uDOF. 

\section*{Acknowledgement}
We would like to acknowledge the computational resources provided by the Aalto Science-IT project.

\end{document}